\documentclass[proof]{pasj00}
\begin{document}
\SetRunningHead{K. Yoshikawa et al.}{Warm/Hot Intergalactic Medium}
\title{Detectability of the Warm/Hot Intergalactic Medium Through
Emission Lines of O{\sc vii} and O{\sc viii}}
\author{
Kohji \textsc{Yoshikawa}\altaffilmark{1,2},
Noriko Y. \textsc{Yamasaki}\altaffilmark{3},
Yasushi \textsc{Suto}\altaffilmark{1,2},\\
Takaya \textsc{Ohashi}\altaffilmark{4},
Kazuhisa \textsc{Mitsuda}\altaffilmark{3},
Yuzuru \textsc{Tawara}\altaffilmark{5},
and
Akihiro \textsc{Furuzawa}\altaffilmark{5}
}

\altaffiltext{1}{Research Center for the Early Universe (RESCEU), School
of Science, \\University of Tokyo, Tokyo 113-0033}
\email{kohji@utap.phys.s.u-tokyo.ac.jp} 

\altaffiltext{2}{Department of Physics, School of Science, University of
Tokyo, Tokyo 113-0033}

\altaffiltext{3}{The Institute of Space and Astronautical Science
(ISAS), \\ 3-1-1 Yoshinodai, Sagamihara, Kanagawa 229-8510}

\altaffiltext{4}{Department of Physics, Tokyo Metropolitan University,
1-1 Minami-Osawa, Hachioji, Tokyo 192-0397}

\altaffiltext{5}{Department of Physics, Nagoya University, 
Furo-cho, Chikusa-ku, Nagoya 464-8602}
\Received{2003/02/25}
\Accepted{}

\maketitle


\begin{abstract}
Most of cosmic baryons predicted by the big-bang nucleosynthesis has
 evaded the direct detection.  Recent numerical simulations indicate
 that approximately 30 to 50 percent of the total baryons in the present
 universe is supposed to take a form of warm/hot intergalactic medium
 (WHIM) whose X-ray continuum emission is very weak.  To identify those
 missing baryons, we consider in detail the detectability of WHIM
 directly through emission lines of O{\sc vii} (561, 568, 574, 665eV)
 and O{\sc viii} (653eV).  For this purpose, we create mock spectra of
 the emission lines of WHIM using a light-cone output of cosmological
 hydrodynamic simulations.  While the predicted fluxes are generally
 below the current detection limit, an unambiguous detection will be
 feasible with a dedicated X-ray satellite mission that we also discuss
 in detail.  Our proposed mission is especially sensitive to the WHIM
 with gas temperature $T=10^{6-7}$K and overdensity $\delta=10-100$ up
 to a redshift of $0.3$ without being significantly contaminated by the
 cosmic X-ray background and the Galactic emissions. Thus such a mission
 provides a unique and important tool to identify a large fraction of
 otherwise elusive baryons in the universe.
\end{abstract}

\section{Introduction}

{\it Missing mass} or dark matter in the universe has been well
recognized as one of the central problems both in observational
cosmology and in astro-particle physics for more than 20
years. Nevertheless the nature of dark matter is hardly understood yet.
{\it Missing baryons}, on the other hand, do not seem to have attracted
much attention until very recently despite the fact that baryons are
clearly the best understood component in the universe.
\citet{Fukugita1998} are one of the first who explicitly pointed out the
importance of studying the missing baryons, or {\it the cosmic baryon
budget}. Subsequent numerical simulations (e.g., \cite{Cen1999a,%
Dave2001}) indeed suggest that approximately 30 to 50 percent of total
baryons at $z=0$ take the form of the warm-hot intergalactic medium
(WHIM) with $10^5 {\rm K}< T < 10^7 {\rm K}$ which has evaded any direct
detection so far. Unfortunately WHIM does not exhibit strong
observational signature, and its detection has been proposed only either
through the O{\sc vi} absorption features in the QSO spectra
\citep{Hellsten1998, Perna1998, Fang2000, Cen2001, Fang2001, Fang2002a}
or the possible contribution to the cosmic X-ray background in the soft
band \citep{Croft2001, Phillips2001}.

\citet{Tripp2000} and \citet{Tripp2001} reported the first detection of
the WHIM at low redshifts ($z\simeq0.2$) through the O{\sc vi}
absorption lines in the spectra of a bright QSO H1821+643 observed by
{\it FUSE} satellite, and they found that the absorbers are
collisionally ionized with the metallicity $Z>0.1Z_{\odot}$ and have the
number density $dN/dz\simeq50$. Using {\it Chandra} observatory,
\citet{Nicastro2002} and \citet{Fang2002b} also detected the absorption
lines of O{\sc vii}, O{\sc viii} and Ne{\sc ix} at $z<0.1$ in the
spectrum of PKS 2155-304. Their spectral analyses indicate that the
number density and temperature of the absorber are $n_{\rm e}\simeq
6\times10^{-6}$ cm$^{-3}$ and $T=5\times10^5$ K, respectively, with the
coherent size being $\sim 3$ Mpc. This may be interpreted as WHIM in the
nearby filamentary structures.  \citet{Simcoe2002} observed the high
redshift ($2.2<z<2.7$) O{\sc vi} absorption line systems in several QSO
spectra. Those are also interpreted as WHIM at high redshifts; the
strong spatial clustering of the absorbers implies that they may indeed
trace the large-scale structure in the universe.

In those attempts to detect WHIM through the absorption line systems,
the observations are quite limited to particular lines-of-sight toward
QSOs. The unbiased exploration of WHIM, however, requires a systematic
survey-type observation, which motivates us to consider a possibility
to identify emission line features of WHIM, in particular, those of
O{\sc vii} (561eV:1s$^2$--1s2s), O{\sc vii} (568eV:1s$^2$--1s2p),
O{\sc vii} (574eV:1s$^2$--1s2p), O{\sc vii} (665eV:1s$^2$--1s3p) and
O{\sc viii} (653eV:1s--2p).  The major obstacle against this attempt
is the Galactic emission; our Galaxy harbors a significant amount of
interstellar gas with similar temperature with WHIM, and the Galactic
emission is typically two orders of magnitude stronger than the
expected WHIM emission \citep{McCammon2002}.  For instance, recently
\citet{Kaastra2003} claimed the detection of O{\sc vii} emission lines
in the outskirts of galaxy clusters with {\it
XMM--Newton}. Unfortunately because of the limited energy resolution
($\simeq 80$ eV) of the mission, it is not easy to distinguish the
WHIM and the Galactic components in a convincing manner. This
indicates the importance of the energy resolution of a detector for a
reliable identification of WHIM.  In fact, we show below that with an
energy resolution of $\Delta E \sim 2$eV which is achievable in near
future, one can distinguish the redshifted emission lines of O{\sc
vii} and O{\sc viii} from other prominent Galactic lines.

The rest of the paper is organized as follows; section 2 summarizes the
outline of our cosmological simulations and the model for metallicity
evolution.  Section 3 discusses the major noise sources and requirements
for the instrument for the successful detection of oxygen emission
lines.  We present the predicted mock spectra assuming fiducial
parameters for our proposed mission in section 4, and consider consider
the detectability of the WHIM in section 5.  Finally section 6 is
devoted to implications and discussion of the present paper.

\section{Modeling Metallicity Evolution of Intergalactic Medium}

\subsection{Cosmological Hydrodynamic Simulation of Intergalactic Medium}

We use the simulation data of \citet{Yoshikawa2001}, which are briefly
described here.  The simulation code is a hybrid of
Particle--Particle--Particle--Mesh (PPPM) Poisson solver and smoothed
particle hydrodynamics (SPH). We employ $128^3$ dark matter particles
and the same number of gas particles within the periodic simulation cube
of $L_{\rm box} =75h^{-1}\mbox{Mpc}$ per side. We adopt a spatially-flat
$\Lambda$CDM universe with $\Omega_0=0.3$, $\Omega_{\rm b}=0.015h^{-2}$,
$\lambda_0=0.7$, $\sigma_8=1.0$, and $h=0.7$, where $\Omega_0$ is the
density parameter, $\Omega_{\rm b}$ is the baryon density parameter,
$\lambda_0$ is the dimensionless cosmological constant, $\sigma_8$ is
the rms density fluctuation top-hat smoothed over a scale of
$8h^{-1}\mbox{Mpc}$, and $h$ is the Hubble constant in units of 100
km/s/Mpc. Thus the mass of each dark matter and gas particle is
$2.4\times 10^{9} M_{\odot}$ and $2.2\times 10^{10} M_{\odot}$,
respectively. The initial condition is generated at redshift $z=36$ and
is evolved up to the present.

The ideal gas equation of state with an adiabatic index $\gamma=5/3$ is
adopted, and the effect of radiative cooling is incorporated using the
cooling rate of \citet{Sutherland1993} adopting a metallicity
[Fe/H]$=-0.5$ We also implement the cold gas decoupling technique for
the phenomenological treatment of the multi-phase nature of baryons; we
decouple radiatively cooled gas particles from the remaining gas
particles and regard the former as stellar components in galaxies, which
provides an approximate prescription of galaxy formation.  The effects
of energy feedback from supernovae and the UV background heating are
ignored in our simulation.

\begin{figure}[tbp]
 \leavevmode
 \begin{center}
  \FigureFile(160mm,160mm){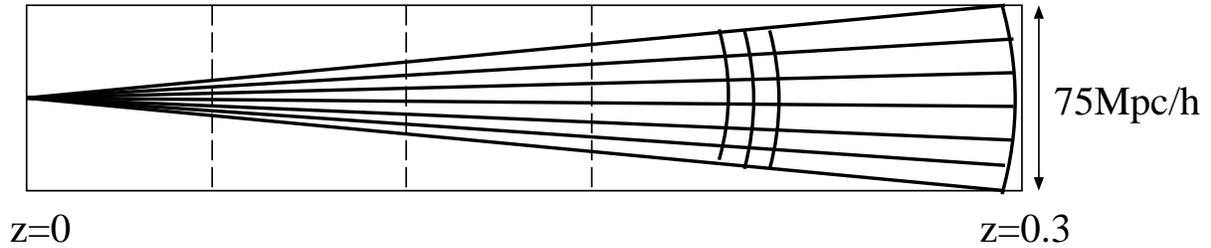} 
 \end{center}
 \caption{A schematic picture of our lightcone output in the comoving
coordinate. The eleven cubic simulation data are stacked along the line
of sight up to $z=0.3$ as described in the text. The surface brightness
are computed on $64\times64$ grids over the $5^\circ \times 5^\circ$
celestial area.  Thus the angular size of the simulated grid is
$5^\circ/64=4.7$ arcmin.  \label{fig:lightcone}}

\end{figure}

All the cosmological observations are carried out over the past
lightcone. In the present study, this is particularly important because
we do not know {\it a priori} the redshifts of the corresponding WHIM
along the line-of-sight. In order to properly model a realistic
observation surveying a $5^\circ \times 5^\circ$ region, we create the
lightcone data as schematically illustrated in
Figure~\ref{fig:lightcone}; we first stack eleven simulation cubes at
different redshifts sampled from $z=0$ to $z=0.3$.  The size of the
survey area corresponds to the simulation boxsize at $z=0.3$; $d_{\rm
A}(z=0.3) \times 5^\circ \approx L_{\rm box}/(1+0.3)$ where $d_{\rm
A}(z)$ is the angular diameter distance toward a redshift $z$.
Specifically we use data at $z=0$, $0$, $0.047$, $0.047$, $0.096$,
$0.096$, $0.148$, $0.148$, $0.202$, $0.202$, and $0.259$ (due to the
limited number of outputs, we sometimes have to use data at the same
redshift twice).  In order to avoid the artificial coherent structure
along the line of sight, we randomly choose a position in the cube and
periodize the particle positions, and also randomly exchange the $x$-,
$y$- and $z$-axes in constructing the light-cone output.

\subsection{Metallicity of the intergalactic medium}

Since the resolution of our cosmological simulations is insufficient to
trace the metallicity evolution, we have to adopt an ad-hoc model for
the chemical evolution of the intergalactic medium (IGM).  Oxygens that
we are interested in here are mainly produced via Type II supernovae in
galaxies, and are generally believed to be ejected into intergalactic
space via the supernova-driven galactic wind and/or other dynamical
processes like mergers and tidal interaction of two galaxies (e.g.,
\cite{Gnedin1997, Nath1997, Gnedin1998, Cen1999a, Cen1999b,%
Ferrara2000}). For definiteness, we consider the following four
possibilities in order to bracket the plausible range of observational
and theoretical uncertainties.

\begin{description}
 \item[model I: ] the metallicity is set to be $Z=0.2Z_{\odot}$
independently of redshifts and the densities of the IGM.
 \item[model II: ] the metallicity is assumed to be spatially uniform
but increase in proportion to the cosmic time $t$:
\begin{equation}
Z=0.2(t/t_0)Z_{\odot},
\end{equation}
where $t_0$ is the age of the universe.
 \item[model III: ] the metallicity is assumed to be correlated with the
local IGM density $\rho_{\rm IGM}({\bf x},t)$ 
relative to the mean baryon density
of the universe $\bar\rho_{\rm b}(t)$. To be specific, we adopt
\begin{equation}
Z=0.005(\rho_{\rm IGM}/\bar{\rho}_{\rm b})^{0.33}Z_{\odot}
\end{equation}
from our fit to the metallicity--density relation at $z=0$ of
the galactic wind driven metal ejection model of
\citet{Aguirre2001b}.
 \item[model IV: ] we assume the similar gas density dependence of the
metallicity as model III but use the fit to the radiation pressure
ejection model of \citet{Aguirre2001b} where the stellar light exerts
radiation pressure on the interstellar dust grains and expels them into
the hosting galactic halos and the ambient IGM.  Actually, we find that
the results of \citet{Aguirre2001b} in this model have the density
dependence of the metal distribution similar to that of model III except
for the overall normalization:
\begin{equation}
   Z=0.02(\rho_{\rm IGM}/\bar{\rho}_{\rm b})^{0.3}Z_{\odot} .
\end{equation}
Note that \citet{Aguirre2001b} concluded that only the radiation
pressure ejection model (model IV) reproduces the typical metallicity in
intracluster medium, while the observed metallicity of Lyman-$\alpha$
clouds at $z\approx 2-3$ may be accounted for by both models III and IV.
\end{description}

\section{Contaminating Sources and Requirements for the Instruments}

In this section, we discuss the contaminating sources in detecting the
emission lines of WHIM and the requirements for the detector and the
telescope to evade such contaminations. 

The major contaminating sources are the cosmic X-ray background (CXB)
and the emission lines of hot gas which resides in the Galactic halo.
According to \citet{McCammon2002}, N{\sc vii} (500eV) and C{\sc vi}
(368eV) are the other prominent lines in the Galactic emission below
oxygen line energies. The O{\sc viii} line of WHIM at $z<0.3$ that we
consider here is free from the confusion with the above two lines, and
thus we focus on the energy range between 500eV and 700eV. 

Thus the detection threshold in this energy range is basically
determined by the intensity of the CXB.  \citet{Miyaji1998} suggested
that the absolute intensity of the CXB below 1 keV has a systematic
error of 20--30 \% in comparison with the {\it ROSAT} and {\it ASCA}
measurements.  Strictly speaking, however, the soft component of the
``CXB'' is supposed to be dominated by the Galactic diffuse emission.
\citet{Kushino2002} found that the {\it ASCA} GIS diffuse background
spectrum after removing the point source contributions consists of a
hard power-law component and a soft component.  More importantly they
found that the intensity of the soft component varies significantly from
field to field (its $1\sigma$ fractional variation amounts to
$52^{+4}_{-5}$\% of the average CXB intensity), and indeed becomes
stronger toward the Galactic Center; the spatial distribution is well
fitted with a finite disk model with a radius of $(1.15\pm0.23)R_{g}$, 
and a scale height of $(0.19\pm0.08)R_{g}$, where $R_{g}$
is the distance to the Galactic center.  This implies that the soft
component (below 1 keV) is dominated by the Galactic diffuse emission.
They estimated the average flux of 6$\times10^{-9}$ erg
cm$^{-2}$s$^{-1}$sr$^{-1}$ between 0.5 and 0.7 keV by averaging 91
fields, corresponding to the flux density of $f_{\rm B} =
3\times10^{-8}$ erg cm$^{-2}$s$^{-1}$sr$^{-1}$keV$^{-1}$.

The primary requirement for the detector of the WHIM emissions is its
high energy resolution to distinguish the Oxygen lines of WHIM at an
intermediate redshift from their Galactic counterparts.  The
observational features of the Galactic emission lines are described in
detail by \citet{McCammon2002}.  They observed $\sim1$ sr sky centered
at $l=90^{\circ}, b=60^{\circ}$ with a resolution of 9 eV, and found the
line intensities of $4.8\pm0.8$ photons cm$^{-2}$s$^{-1}$sr$^{-1}$
($4.4\pm0.7\times10^{-9}$ erg cm$^{-2}$s$^{-1}$ sr$^{-1}$ ) for the
O{\sc vii} triplet and of $1.6\pm0.4$ photons cm$^{-2}$s$^{-1}$sr$^{-1}$
($1.7\pm0.4\times10^{-9}$ erg cm$^{-2}$s$^{-1}$ sr$^{-1}$ ) for the 653
eV O{\sc viii} line.  With a better energy resolution ($\Delta E <$6
eV), the lines of $\delta z \sim 0.01$ can be separated from the
Galactic components. In addition, the separation of O{\sc vii} triplets
may lead to the estimation of the temperature of WHIM from their line
ratios.

Another important requirement is the high efficiency of the throughput
of the telescope. Since the emission lines of WHIM are generally weak,
we need large $S_{\rm eff}\Omega$ for their successful detection, where
$S_{\rm eff}$ is the effective area and $\Omega_{\rm tot}$ is the total
field-of-view of detectors. To realize a large mirror with $S_{\rm
eff}\Omega_{\rm tot} >$ 100cm$^{2}$deg$^{2}$, we are currently pursuing
the possibility of 4-stage X-ray telescopes (Furuzawa et al. 2003, in
preparation). In contrast to the standard 2-stage X-ray mirrors, 4-stage
telescopes will be able to achieve a shorter focal length and a wider
field-of-view.  The short focal length makes it possible to construct a
smaller detector with the same field-of-view. This is essential for the
X-ray micro-calorimeters to simultaneously achieve a small heat
capacitance and a good energy resolution. We simulated a nested 4-stage
telescope with a diameter of 60 cm and a focal length of 70 cm by a
ray-tracing program, and found that the throughput at 0.7 keV is $S_{\rm
eff}\Omega_{\rm tot}= 220 \eta$ cm$^{2}$deg$^{2}$ with $\eta$ being the
overall efficiency of the detector (Furuzawa et al. 2003, in
preparation).

Consider a detector with the effective area $S_{\rm eff}$, the
field-of-view $\Omega$, and the energy resolution $\Delta E$ at the line
energy $E$.  Then the signal-to-noise ratio of emission lines is written
as
\begin{eqnarray}
 \label{eq:detection_limit}
 (S/N)^2 &=& \frac{\displaystyle 
\left(\frac{\bar f_{\rm em}S_{\rm eff}\Omega T_{\rm exp}}{E}\right)^2}
{\displaystyle 
\left(\frac{f_{\rm B}\Delta E S_{\rm eff}\Omega T_{\rm exp}}{E}\right) 
+ \left(\frac{\bar f_{\rm em}S_{\rm eff}\Omega T_{\rm exp}}{E}\right)} 
\cr
&=& \frac{\bar  f_{\rm em}^2 S_{\rm eff}\Omega T_{\rm exp}}
{(f_{\rm B}\Delta E + \bar f_{\rm em})E}, 
\end{eqnarray}
where $\bar f_{\rm em}$ is the flux of emission lines averaged over the
given size of pixels on the detector, and $T_{\rm exp}$ is the exposure
time. The denominator in the above expression indicates the noise level
contributed from the CXB and the Poisson noise of emission from WHIM in
terms of the corresponding number of photons.  Figure~\ref{fig:limit}
shows the signal-to-noise ratio of O{\sc viii} 653 eV emission lines for
various values of $S_{\rm eff}\Omega T_{\rm exp}$ as a function of a
flux of the lines, where we adopt $\Delta E=2$eV as a nominal energy
resolution of our fiducial detector. This indicates that in the case of
$S_{\rm eff}\Omega T_{\rm exp}=10^7$ cm$^2$ deg$^2$ sec and $10^8$
cm$^2$ deg$^2$ sec, one can detect emission lines with fluxes greater
than $7\times10^{-11}$ erg cm$^{-2}$ sec$^{-1}$ sr$^{-1}$ and
$1.5\times10^{-11}$ erg cm$^{-2}$ sec$^{-1}$ sr$^{-1}$ with sufficient
($S/N>10$) statistical significance, respectively.

Actually, the emission from hot intra-cluster medium (ICM) in galaxy
clusters can be also an important contaminating source. For the central
regions of galaxy clusters, the emission from ICM is predominant, and
thus the identification of O{\sc vii} and O{\sc viii} lines becomes
difficult at directions toward such regions. Therefore, the detection
limit inferred from Equation~(\ref{eq:detection_limit}) may be
interpreted as a lower limit.  This contamination, however, is important
only in central regions of galaxy clusters. Actually, as we will see
below, the emission lines above the detection threshold inferred from
Equation~(\ref{eq:detection_limit}) generally produce detectable
spectral signatures.

Considering the transmission of the entrance window and the quantum
efficiency of the detector, a realistic value of $\eta$ may be $\sim$
0.5.  In what follows, therefore, we assume the detector throughput of
$S_{\rm eff} \Omega_{\rm tot}=100 \mbox{ cm$^2$ deg$^2$}$, and the
spectroscopic energy resolution of $\Delta E=2 \mbox{ eV}$. The
field-of-view $\Omega$ may be chosen somewhere in a range of $0.5^\circ$
to $1^\circ$, and is filled with a detector array of $32\times32$
pixels.  This specification is compared with the on-going/planned X-ray
missions in Table~\ref{tab:detector}, where the expected sensitivities
for the emission lines of current and proposed X-ray satellite missions
are summarized. The first two are the current working missions; {\it
Chandra} \citep{wei00} has a good spatial resolution of $1''.5$ (FWHM)
and {\it XMM-Newton} \citep{str01} has a large effective area of 1200
cm$^{2}$.  While {\it Chandra} has the transmission grating
spectrometers HETG and LETG, and {\it XMM-Newton} has the reflection
grating spectrometer, RGS, the energy resolution of those grating
instruments for diffuse emission lines is worse than those of X-ray
CCDs. Table~\ref{tab:detector} lists the sensitivities of X-ray CCD
detectors, ACIS-S3 for {\it Chandra} and EPIC-pn for {\it XMM-Newton}.
Astro-E II (http://www.astro.isas.ac.jp/astroe) is a Japan-USA mission
which will be launched in 2005.  The XRS detector will be the first
X-ray micro-calorimeter in orbit and have an energy resolution of 6 eV.
Due to the small effective area and the field-of-view, however, the
sensitivity of the XRS is worse than that of XIS consisting of four
X-ray CCDs.  The {\it Constellation-X}
(http://constellation.gsfc.nasa.gov) is a US mission of high throughput
consisting of four space-crafts.  The flux limit is estimated on the
basis of the current design of SXT and TES (transition-edge sensor)
calorimeters with 2 eV energy resolution. The {\it XEUS}
(http://astro.estec.esa.nl/XEUS/) will have a huge X-ray mirror of 6
m$^{2}$ and TES calorimeters.  Since those missions are all designed as
X-ray observatories for general purposes, both a large effective area
and a good spatial resolution are preferred and the field-of-view has to
be necessarily smaller.  On the other hand, the nested 4-stage telescope
that we propose here is feasible even with a small satellite, and still
the sensitivity for the diffuse X-ray line emission exceeds those of the
larger missions. In this sense, it has important and complementary
advantages over other current/future X-ray missions.

\section{Mock Observation of Oxygen Emission Lines from the Simulated
 Warm/Hot IGM}

\subsection{Calculation of {\sc Ovii} and {\sc Oviii} Emission}

We divide the entire lightcone data, which we describe in section 2,
into $64\times 64$ square grids on the celestial plane and 128 grids
equally spaced along the redshift direction (see
Fig.~\ref{fig:lightcone}). Thus we have $64\times 64 \times 128$ cells
in total.  For the $J$-th cell, we compute the surface brightness $S_J$
due to the emission lines of O{\sc vii} and O{\sc viii} as
\begin{equation}
 \label{eq:surface_brightness}
 S_J = \sum_i  \frac{\rho_i m_i}{4\pi(1+z_i)^4\,\Delta A_i}
  \left(\frac{X}{m_{\rm p}}\right)^2\,f_{{\rm e},i}^2 \,\epsilon(T_i,Z_i), 
\end{equation}
\begin{equation}
 \Delta A_i \equiv d_{\rm A}^2(z_i) \Delta \omega ,
\end{equation}
where the summation is over the gas particles (labelled $i$) within the
$J$-th cell.  In the above expression, $X$ is the hydrogen mass fraction
(we adopt $0.755$), $m_{\rm p}$ is the proton mass, $\epsilon(T,Z)$ is
the oxygen line emissivities in units of the power input normalized to
the electron densities as defined by \citet{Mewe1985}, $\Delta
\omega=(5^\circ/64)^2$ is the solid angle of each cell, and $f_{{\rm
e},i}$, $z_i$, $\rho_{i}$, and $m_i$ denote the electron number fraction
relative to the hydrogen, redshift, mass density and mass of the $i$-th
gas particle, respectively.  

In the present study, we assume the collisional ionization equilibrium
in computing $f_{\rm e}$ and $\epsilon$.  Strictly speaking, however,
this assumption may not be completely valid in most regions of the WHIM.
Adopting the typical WHIM parameters ($n_{\rm e}=1\times10^{-5}$
cm$^{-3}$ and $T=10^6$K), for instance, the recombination timescale of
H{\sc ii} $\simeq 3\times10^{11}$ yr significantly exceeds that of of
the collisional ionization ($\simeq 10^5$ yr).  Nevertheless we adopt
the collisional ionization equilibrium for simplicity as in the case of
all the previous simulation studies of the absorption lines due to the
WHIM. We plan to discuss the non-equilibrium effect elsewhere.

\begin{figure}[tbp]
 \leavevmode
 \begin{center}
  \FigureFile(160mm,160mm){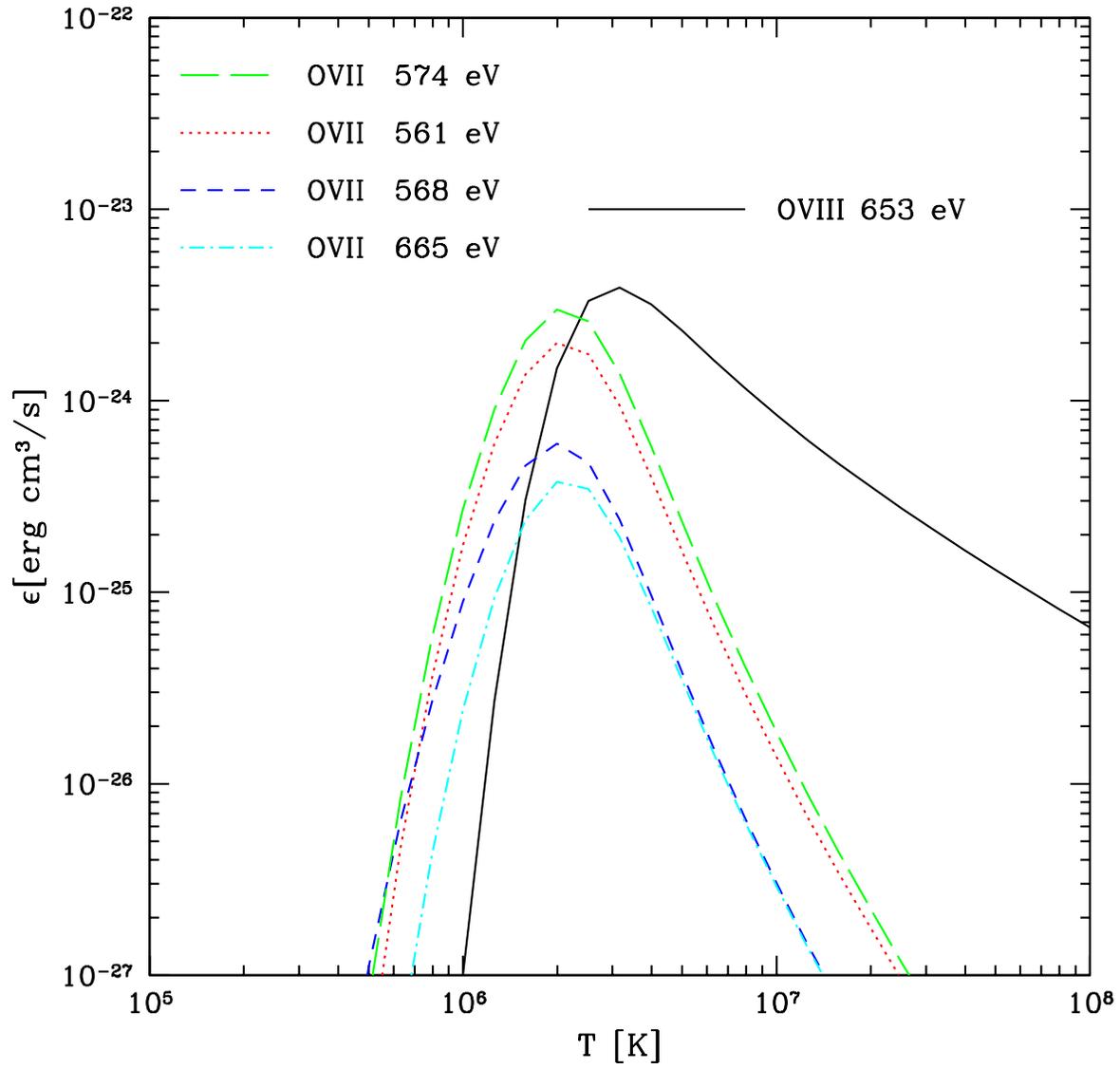}
 \end{center}
  \caption{Emissivity of O{\sc vii} and O{\sc viii} lines in collisional
  ionization equilibrium. \label{fig:emissivity}}
\end{figure}

Figure~\ref{fig:emissivity} shows the emissivities of O{\sc vii} and
O{\sc viii} lines as a function of gas temperature, where we set the
metallicity as $Z=Z_{\odot}$, from the table in the SPEX ver 1.10
(http://www.rhea.sron.nl/divisions/hea/spex).  The intensity of O{\sc
vii} emission lines, a triplet of 574 eV, 581 eV and 588 eV, is peaked
around at $T=2\times10^{6}$ K, and the O{\sc viii} 653 eV line becomes
dominant at $T>3\times10^{6}$ K.

Figure~\ref{fig:OVIII_map} shows the total intensity maps of O{\sc vii}
and O{\sc viii} emission lines of WHIM ({\it left panels}) and the
bolometric X-ray emission ({\it right panels}) integrated over the
redshift range of $0<z<0.3$ ({\it upper panels}), $0.03<z<0.04$ ({\it
middle panels}), $0.09<z<0.11$ ({\it lower panels}). The six squares
labelled A to F indicate the regions whose mock spectra will be
presented in the next section.

\subsection{Calculation of Simulated Soft X-Ray Spectra}

For each of $64\times64$ cells on the celestial plane, we also compute
the spectrum of WHIM and/or ICM at soft X-ray energy band ranging from
450eV to 700eV. The flux intensity at an energy range from $E$ to
$E+\Delta E$ can be calculated as a superposition of spectra for SPH
particles by
\begin{eqnarray}
 \label{eq:spectrum}
 F(E,E+\Delta E) &=& 
\sum_i \frac{\rho_im_i}{4\pi(1+z_i)^4\Delta A_i}
\left(\frac{X}{m_{\rm p}}\right)^2 f_{{\rm e},i}^2 \cr
&\times& \int_{E(1+z_i)}^{(E+\Delta E)(1+z_i)}P(E^\prime,T_i,Z_i)dE^{\prime},
\end{eqnarray}
where the summation is over the SPH particles and $P(E,T,Z)$ is the
template spectrum for temperature $T$ and metallicity $Z$. The set of
template spectra $P(E,T,Z)$ is created using SPEX software package and
is tabulated for temperature range $10^5 \mbox{K} < T < 10^8 \mbox{K}$
and metallicity range $10^{-3}Z_{\odot} < Z < Z_{\odot}$. Here, we
assume again that all the baryons are under collisional ionization
equilibrium in this temperature range.  Figure~\ref{fig:template} shows
the template spectra of collisionally ionized plasma with temperature
$T=10^6$ K, $10^{6.5}$ K, and $10^{7}$ K, and with metallicity
$Z=Z_{\odot}$ and $Z=0.1Z_{\odot}$. At a lower temperature, $T=10^6$ K,
we have strong emission lines of the O{\sc vii} triplets ($E=561, 568,
574$ eV). On the other hand, at $T=10^{6.5}$ K and $10^{7}$ K, O{\sc
viii} line at $E=653$ eV and many Fe\,{\sc xvii} lines at $E>700$ eV
emerge.

Using the template spectra (Fig.~\ref{fig:template}) and the proposed
detector specification, we construct mock emission spectra of WHIM over
the lightcone data as follows.  First we compute the total photon
numbers on each $(5^\circ/64)^2$ grid due to the three components; the
WHIM emission spectra according to Equation~(\ref{eq:spectrum}), the CXB
contribution adopting $f_{\rm CXB}$=6$\times10^{-9}$ erg
cm$^{-2}$s$^{-1}$sr$^{-1}$ between 0.5 and 0.7 keV and a photon index
$\Gamma$=6 \citep{Kushino2002}, and the Galactic emission lines of
Oxygen and Nitrogen as observed by \citet{McCammon2002}.  In practice,
we compute the numbers of photon per each spectroscopic bin assuming a
given exposure time. Then we assign the additional photon number
fluctuations per each bin according to the corresponding Poisson
statistics.  The energy spectrum of photons is convolved with the
response function of a detector which we approximate by a Gaussian with
a FWHM of 2eV.

The upper panel of Figure~\ref{fig:spec_subtracted} shows an example of
the simulated composite spectrum, which includes the contribution from
WHIM, CXB, the Galactic emission lines, with exposure time $T_{\rm
exp}=3\times10^5$ sec for the region A (0.88 deg$^2$) in
Fig.~\ref{fig:OVIII_map}. Strong lines in the upper panel correspond to
the Galactic emission lines of N{\sc vii} at 498 eV, O{\sc vii} at
561eV, 568 eV, 574 eV, and 665 eV, and O{\sc viii} at 653 eV.  Clearly
the separation of the Galactic component from the WHIM emission lines is
the most essential.  In order to mimic the realistic separation
procedure, we construct an independent realization of spectra which
consists purely of the CXB and Galactic emission lines but using the
different sets of random numbers in adding the Poisson fluctuations in
each bin.  Then the latter spectra are subtracted from the mock spectra
(WHIM + CXB and Galactic emissions), which yields a residual mock
spectrum purely for the WHIM ({\it dots with Poisson error bars in lower
panel}).  For comparison, we plot the noiseless WHIM spectrum from
simulation in solid line.  The emission lines with labels in the lower
panel indicate O{\sc viii} and O{\sc vii} triplet lines whose surface
brightness exceeds our nominal limiting flux (see the top-right panel of
Figure~\ref{fig:wide_los}). The O{\sc viii} emission lines from the WHIM
at $z=0.04-0.23$ can be indeed observed as the residual prominent
emission lines between $E=500-630$ eV, though they are also contaminated
by other emission lines. Thus this plot indicates that the emission
lines exceeding the residual photon number $\sim 100$ counts/bin are
detectable with a three-day exposure.

In reality, the continuum spectra from CXB, Galactic diffuse, and WHIM
would be hard to be separated. As indicated in the above idealistic
example, however, the continuum level of the WHIM emission is just
within the fluctuation of the CXB and the Galactic emission. Thus the
uncertainty of the continuum level does not affect the identification of
the emission lines using the proper line fitting procedure.

The absorption due to the interstellar medium of the Galaxy is another
possible source for systematic errors in detecting the WHIM emission. At
a photon energy of $E=0.6$keV, the dominant absorption is due to the
Galactic oxygen as well as other metals. The oxygen abundance is usually
assumed to be in proportion to the Galactic neutral hydrogen. So one can
use the observed column density of neutral hydrogen $N_{\rm H}$ in
inferring the degree of the absorption.  Figure 1 of
\citep{Morrison1983} indicates that the absorption cross section around
$E=0.6$keV is $\sigma \approx 7.4\times10^{-22}$cm$^2$ for 1 solar
metallicity.  So the regions of the sky with $N_{\rm H} = 3 \times
10^{20}$ cm$^{-2}$ and $10^{20}$ cm$^{-2}$ decrease the soft X-ray
intensity around $E=0.6$keV by 20\% and 7\%, respectively.  In other
words, if we preferentially select survey fields with $N_{\rm H} <
10^{20}$ cm$^{-2}$, the absorption is less than 10\%. Therefore even
taking account of the uncertainty of the metal abundance and $N_{\rm
H}$, we expect that the absorption effect may be corrected within a few
percent accuracy.

\section{Results}

We illustrate results corresponding to two different types of
observational strategies.  One is a relatively shallow blank survey
which attempts to detect the WHIM over the large fields in an unbiased
manner. The other is a deeper survey which observes the pre-selected
areas, for instance, the regions surrounding X-ray luminous clusters to
search for the connecting filamentary WHIM from the clusters.  The
detector that we have roughly in mind here has an array of $32\times32$
pixels over the total field-of-view of $\Omega_{\rm tot}=(0.5-1.0)$
deg$^2$, which has a better angular resolution than that of our
simulation grids ($5^\circ/64=4.7$ arcmin). The higher angular
resolution is in any case useful in removing the point sources, etc.  In
reality, however, a reasonably low detection threshold, i.e., $S_{\rm
eff}\Omega T_{\rm exp}> 10^7$ cm$^2$ deg$^2$ sec discussed in the
previous subsection, can be achieved only by merging many pixels at the
expense of the angular resolution depending on the adopted value of
$T_{\rm exp}$. Here, we assume that the total field-of-view is
$\Omega_{\rm tot}=0.88$ deg$^2$ which corresponds to $12\times12$
simulated grids.

\begin{figure}[tbp]
 \begin{center}
  \FigureFile(65mm,65mm){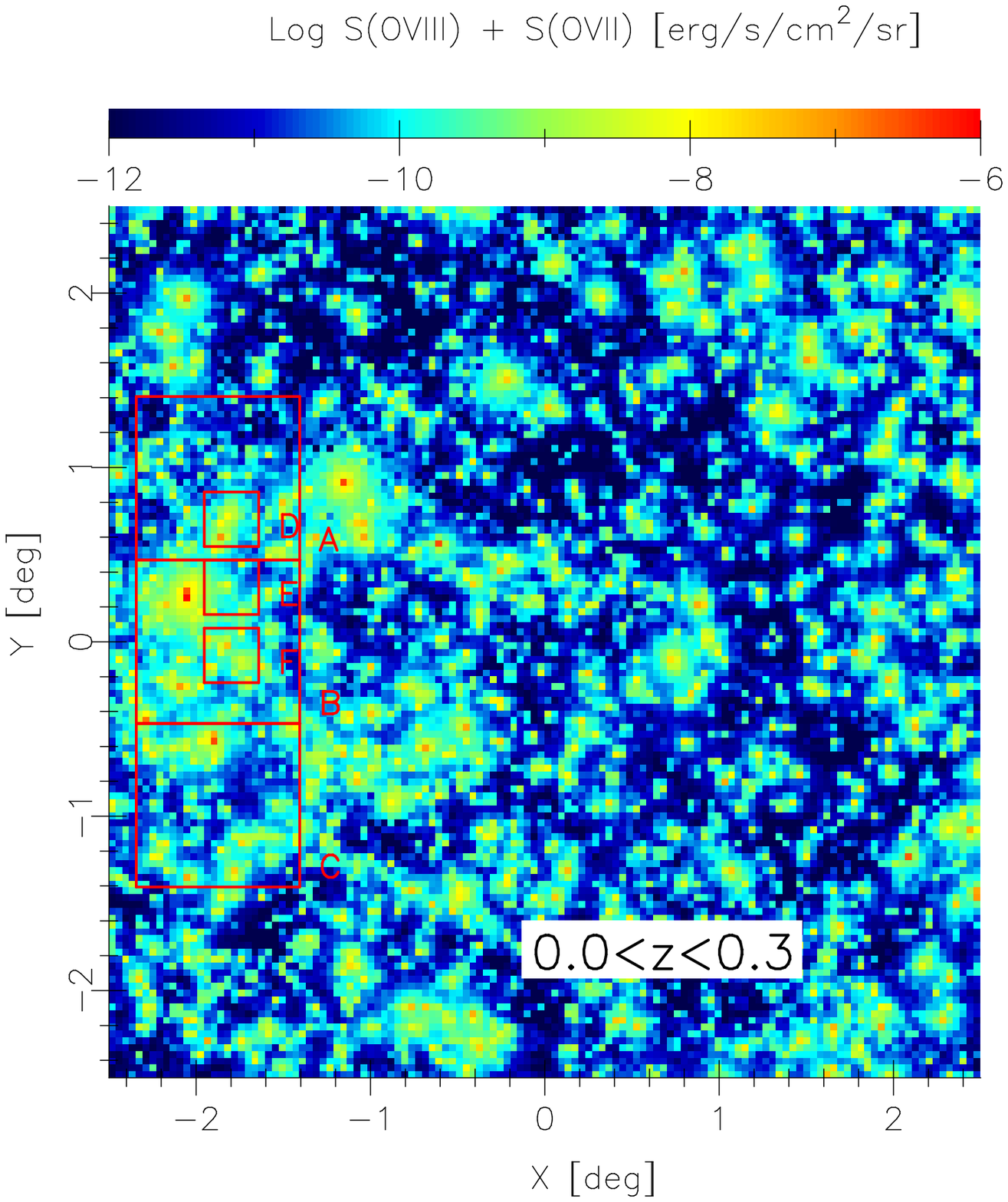} 
  \FigureFile(65mm,65mm){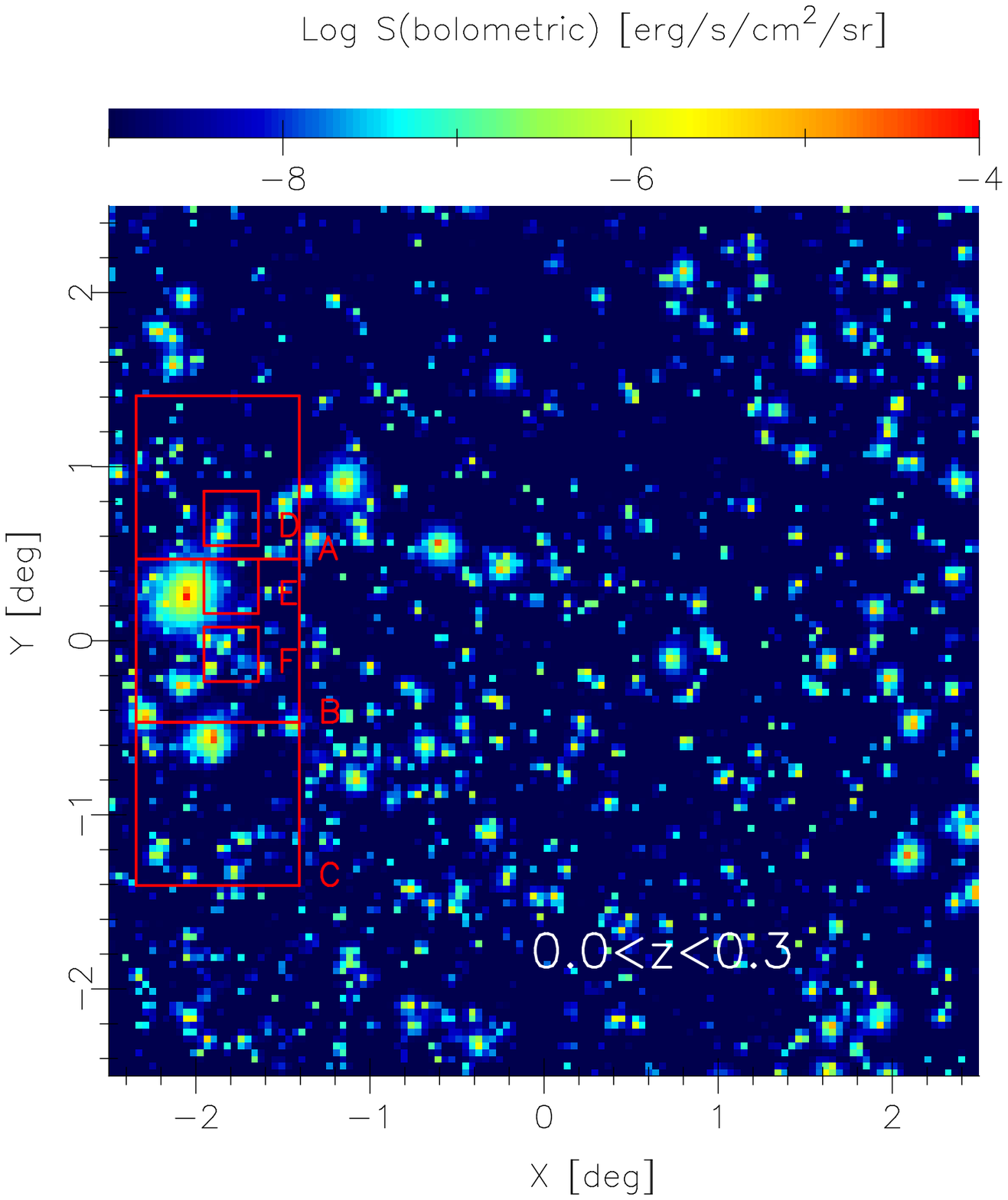} 
  \FigureFile(65mm,65mm){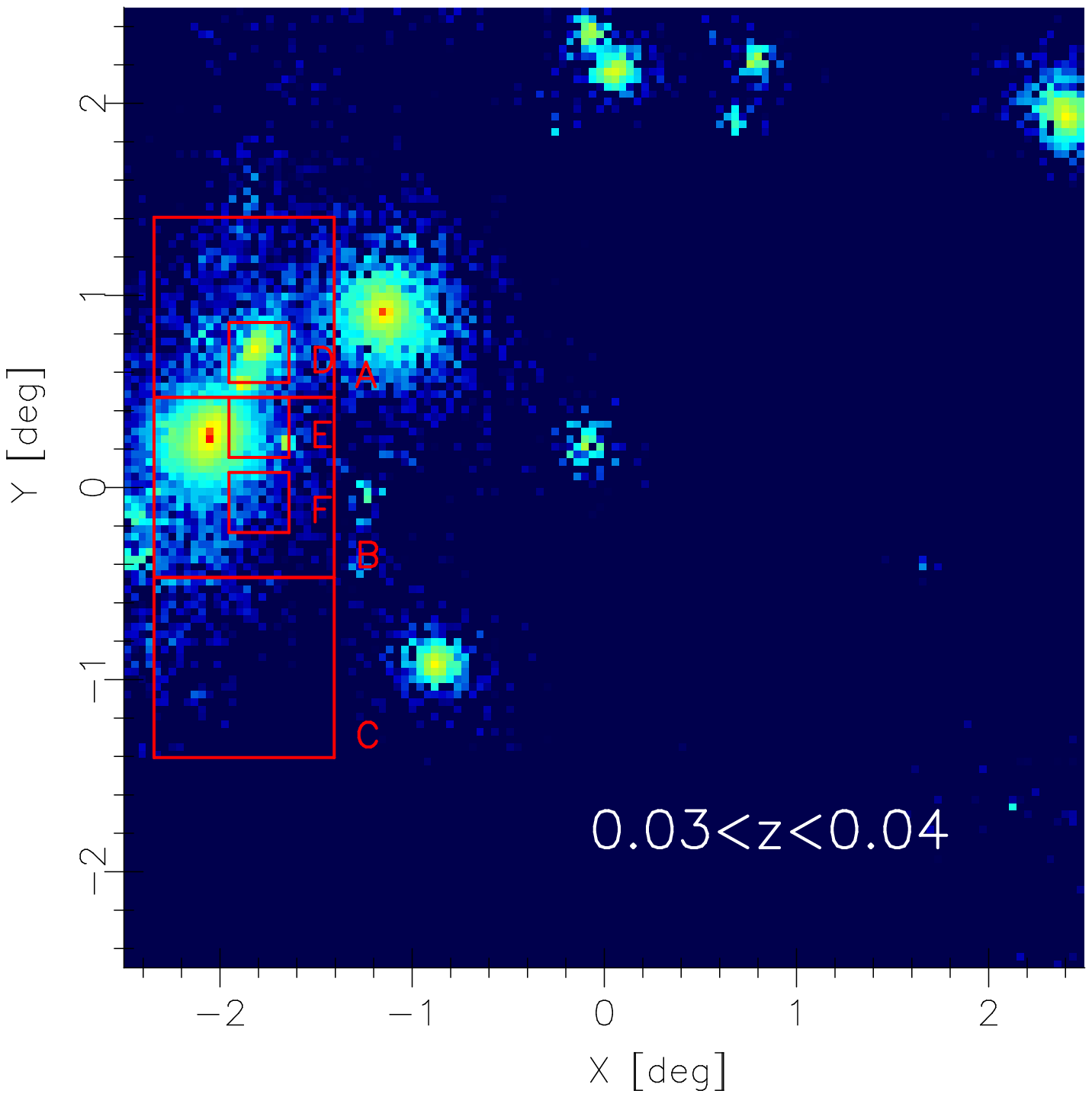}
  \FigureFile(65mm,65mm){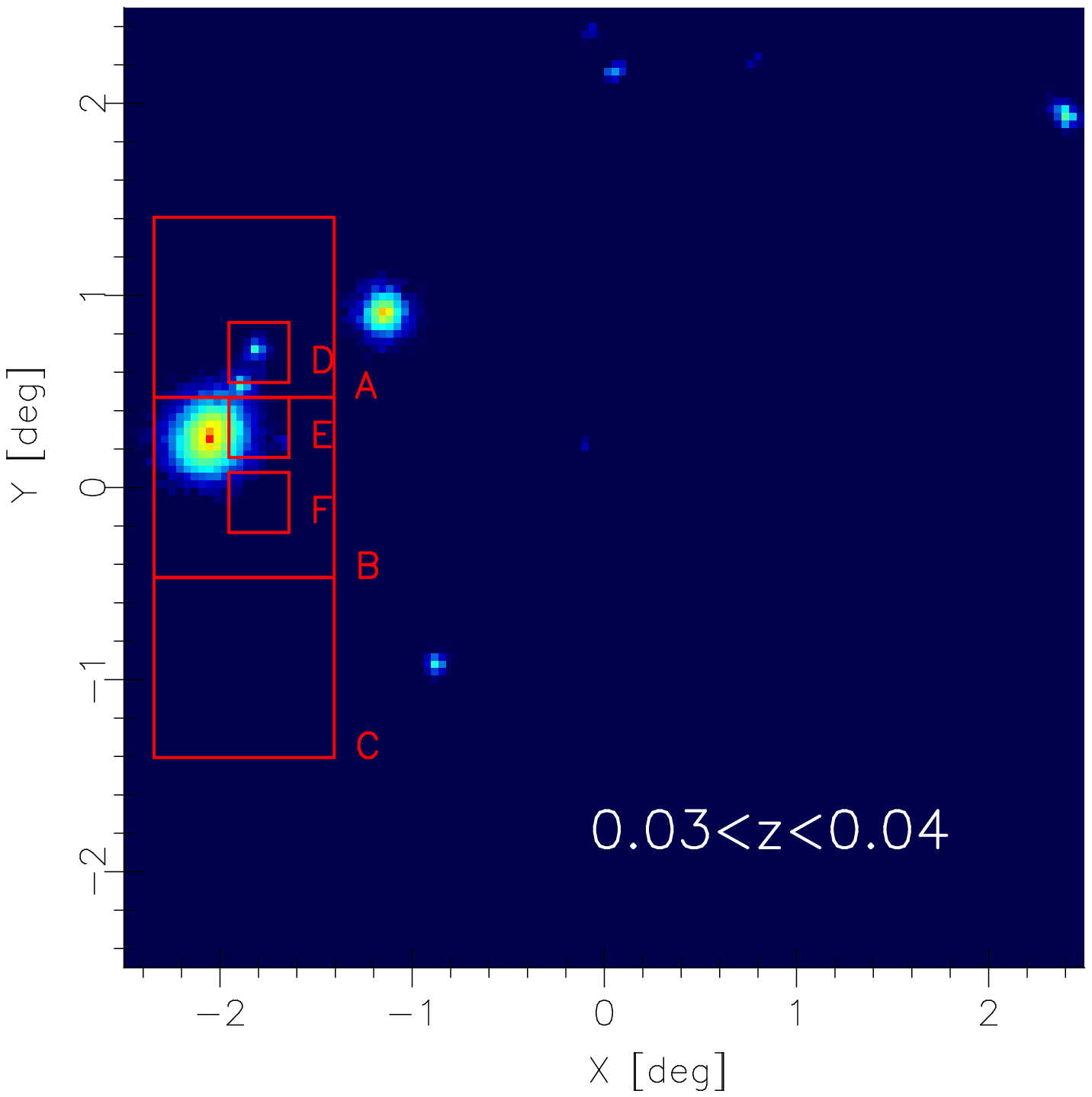} 
  \FigureFile(65mm,65mm){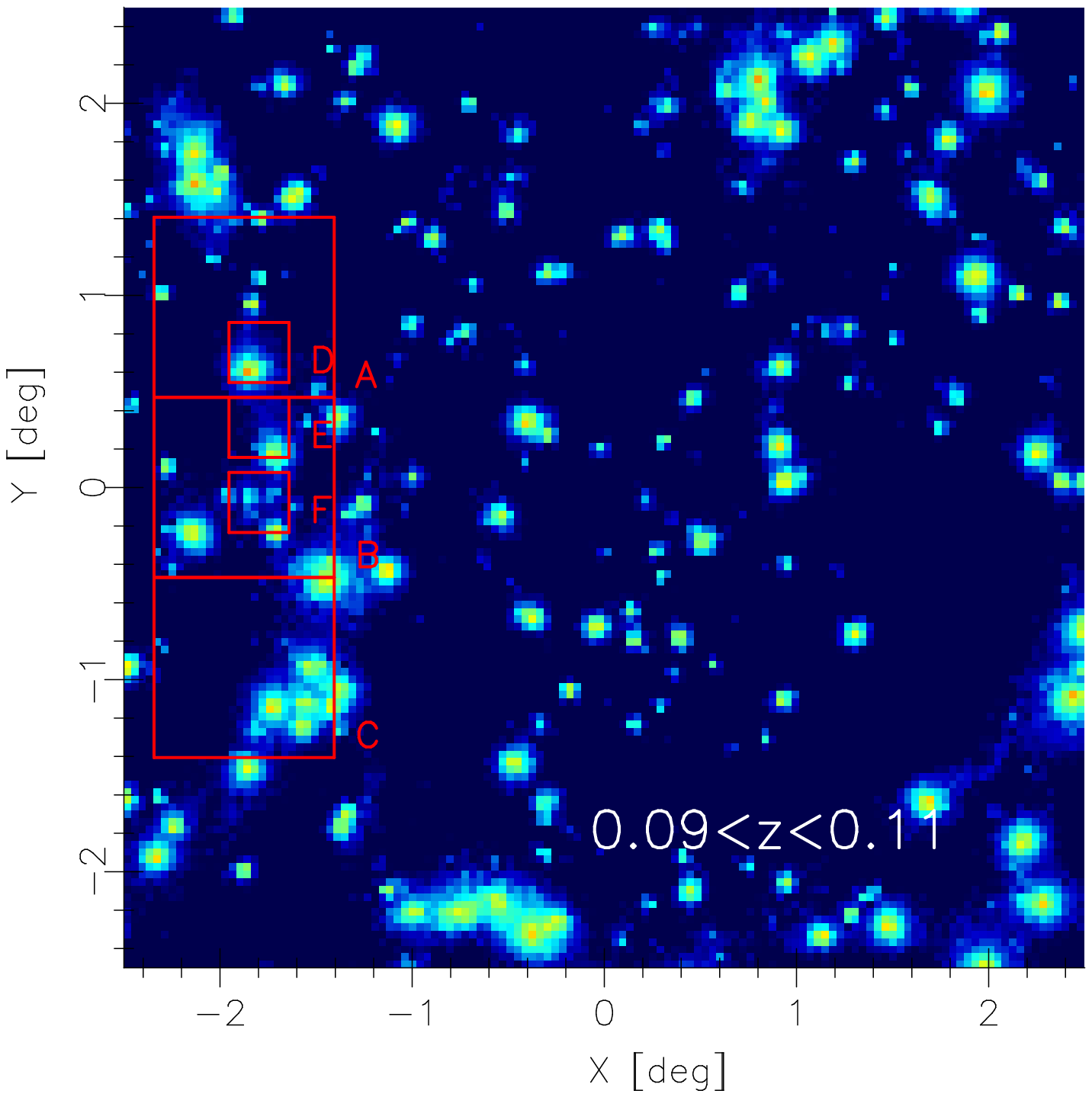}
  \FigureFile(65mm,65mm){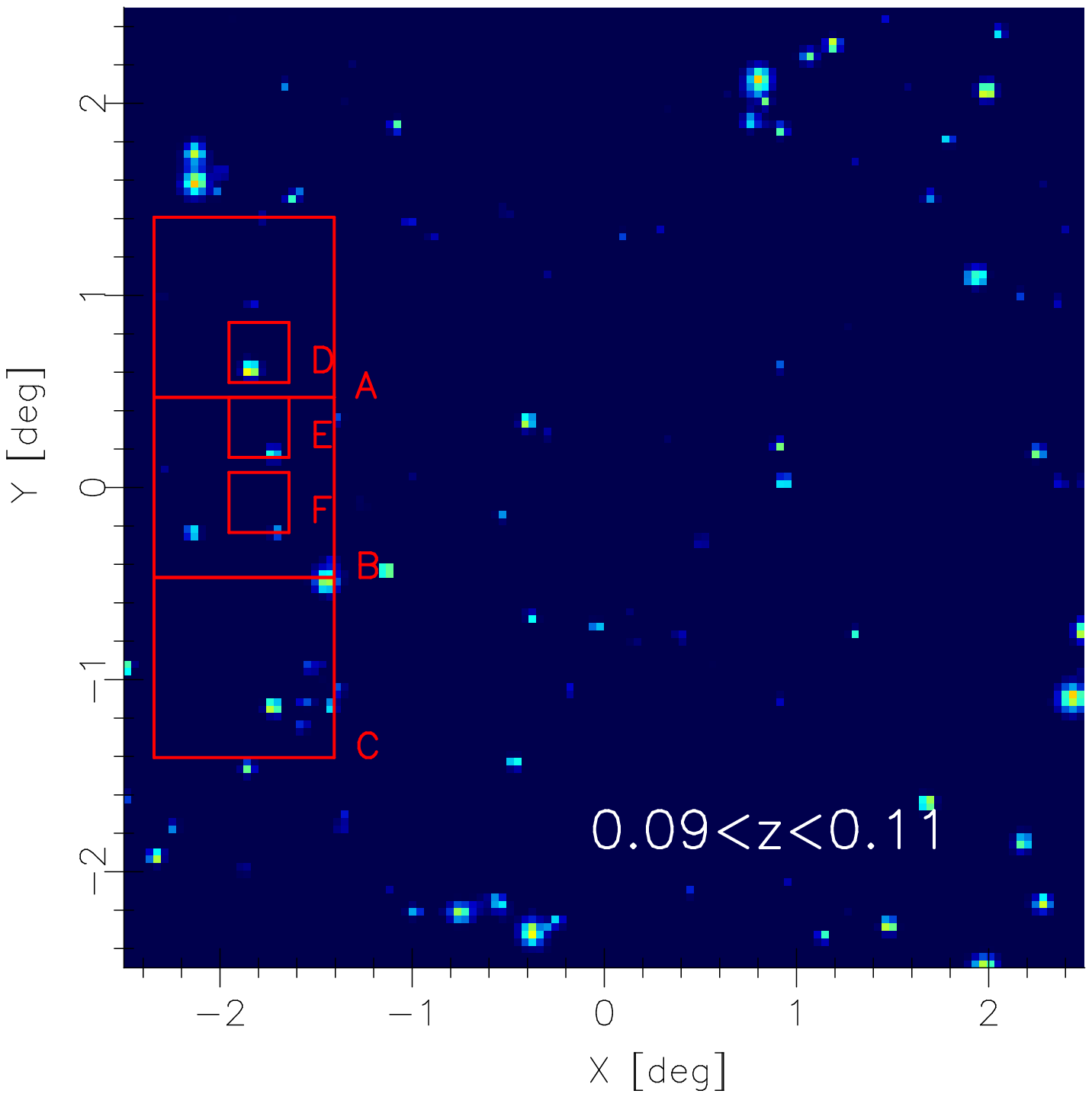}
 \end{center}
 \vspace{-0.5cm}
 \caption{The map of O{\sc viii} and O{\sc vii} emission ({\it left
 panels}) and X-ray bolometric emission ({\it right panels}) from $0.0 <
 z < 0.3$ ({\it upper panels}), from $0.03 < z < 0.04$ ({\it middle
 panels}), and from $0.09 < z < 0.11$ ({\it lower panels}). Squares
 indicate regions of which the simulated spectra in
 Figure~\ref{fig:wide_los} and \ref{fig:deep_los} are
 calculated. \label{fig:OVIII_map}}

\end{figure}

\begin{figure}[h]
 \begin{center}
   \FigureFile(130mm,130mm){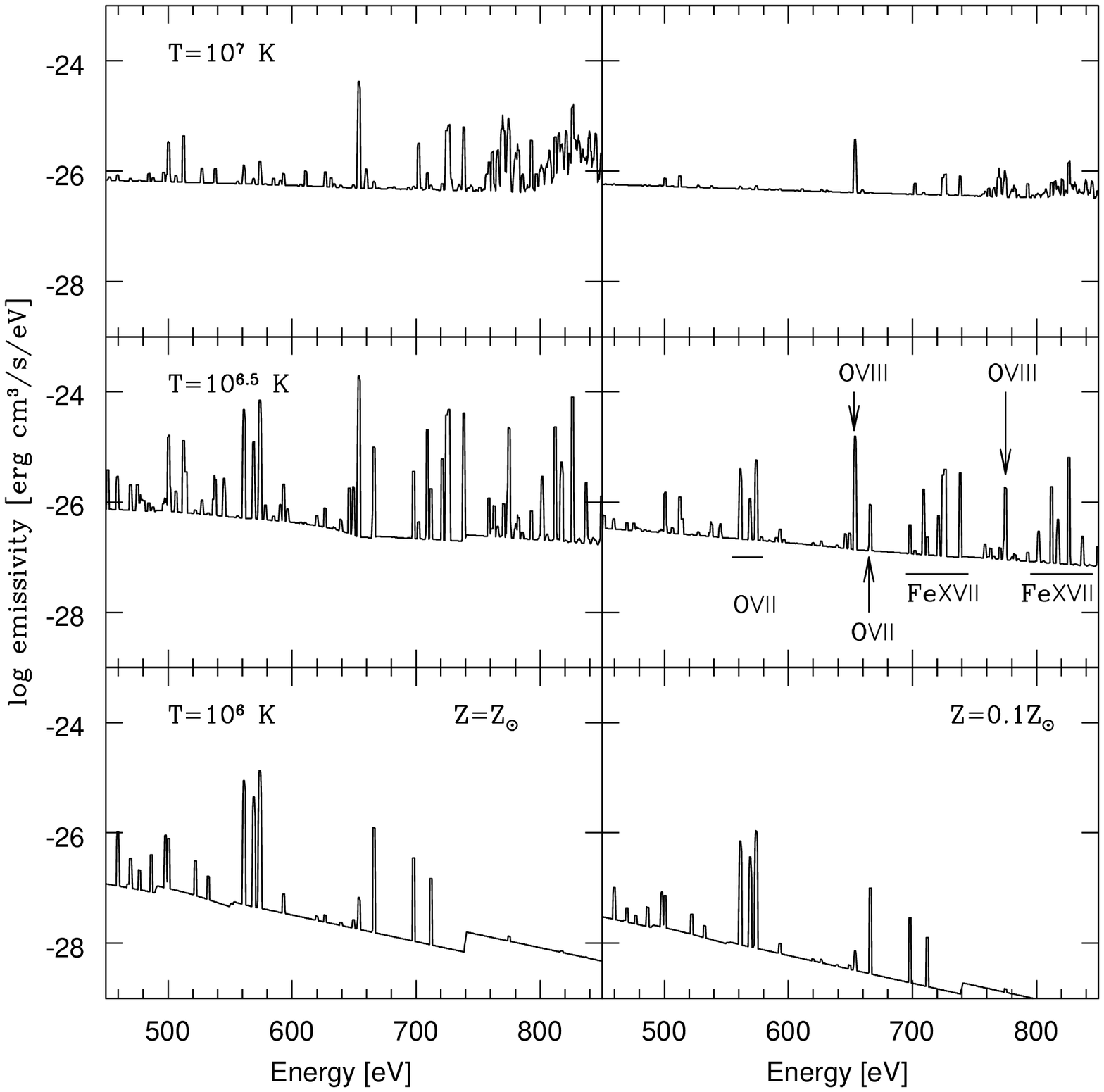}
 \end{center}
 \vspace{-0.5cm}
 \caption{Template spectra of collisionally ionized plasma with
 temperature $T=10^6$ K ({\it lower panels}), $10^{6.5}$ K ({\it middle
 panels}), and $10^7$ K ({\it upper panels}). Spectra for metallicity
 $Z=Z_{\odot}$ and $Z=0.1Z_{\odot}$ are shown in the left and right
 panels, respectively. \label{fig:template}}

 \begin{center}
  \FigureFile(120mm,120mm){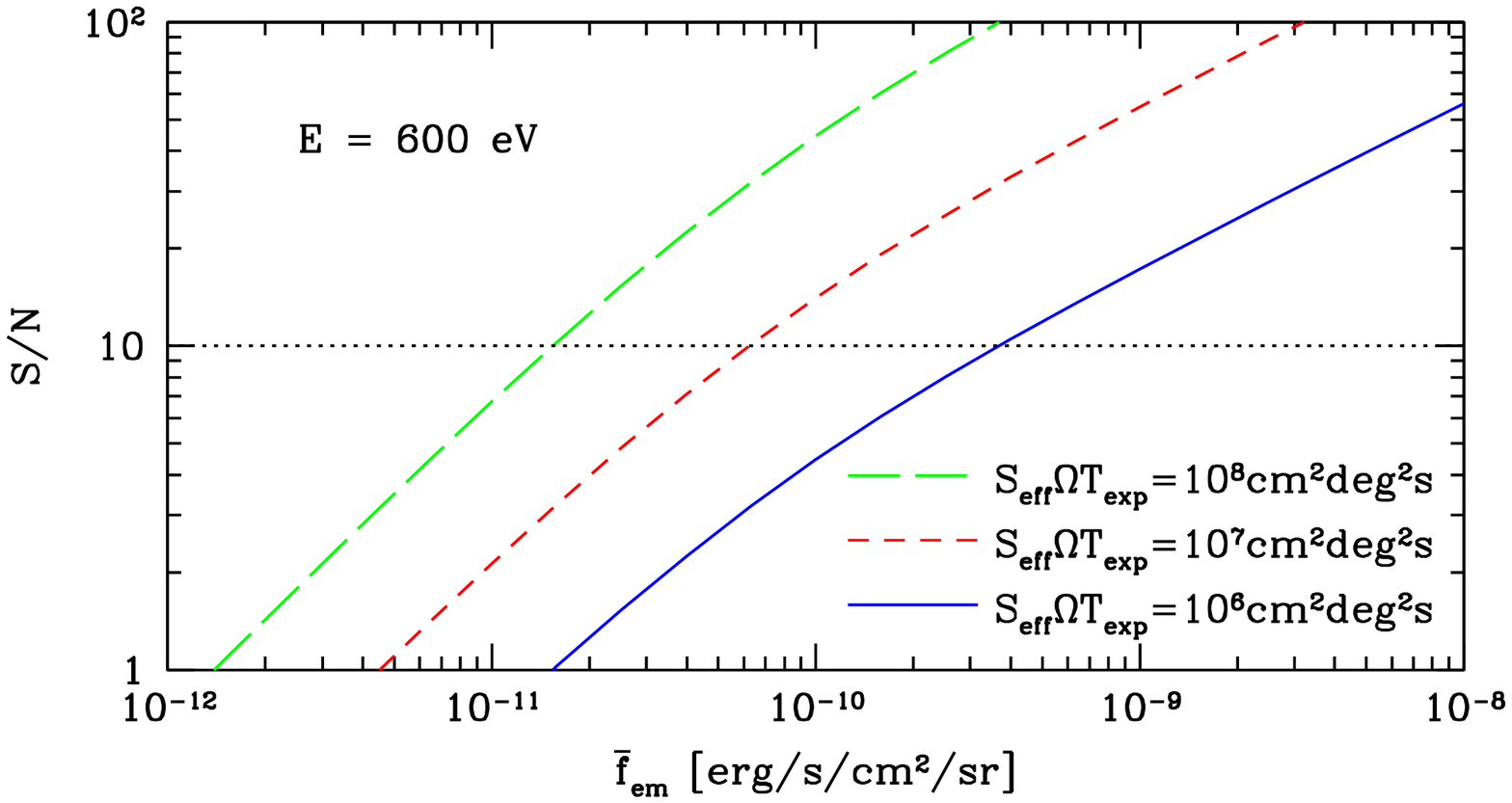}
 \end{center}
 \caption{Signal-to-Noise ratio of O{\sc viii} emission lines embedded
 in the CXB for various values of $S_{\rm eff}\Omega T_{\rm
 exp}$.\label{fig:limit}}
\end{figure}

\begin{figure}[tbp]
 \leavevmode
 \begin{center}
  \FigureFile(160mm,160mm){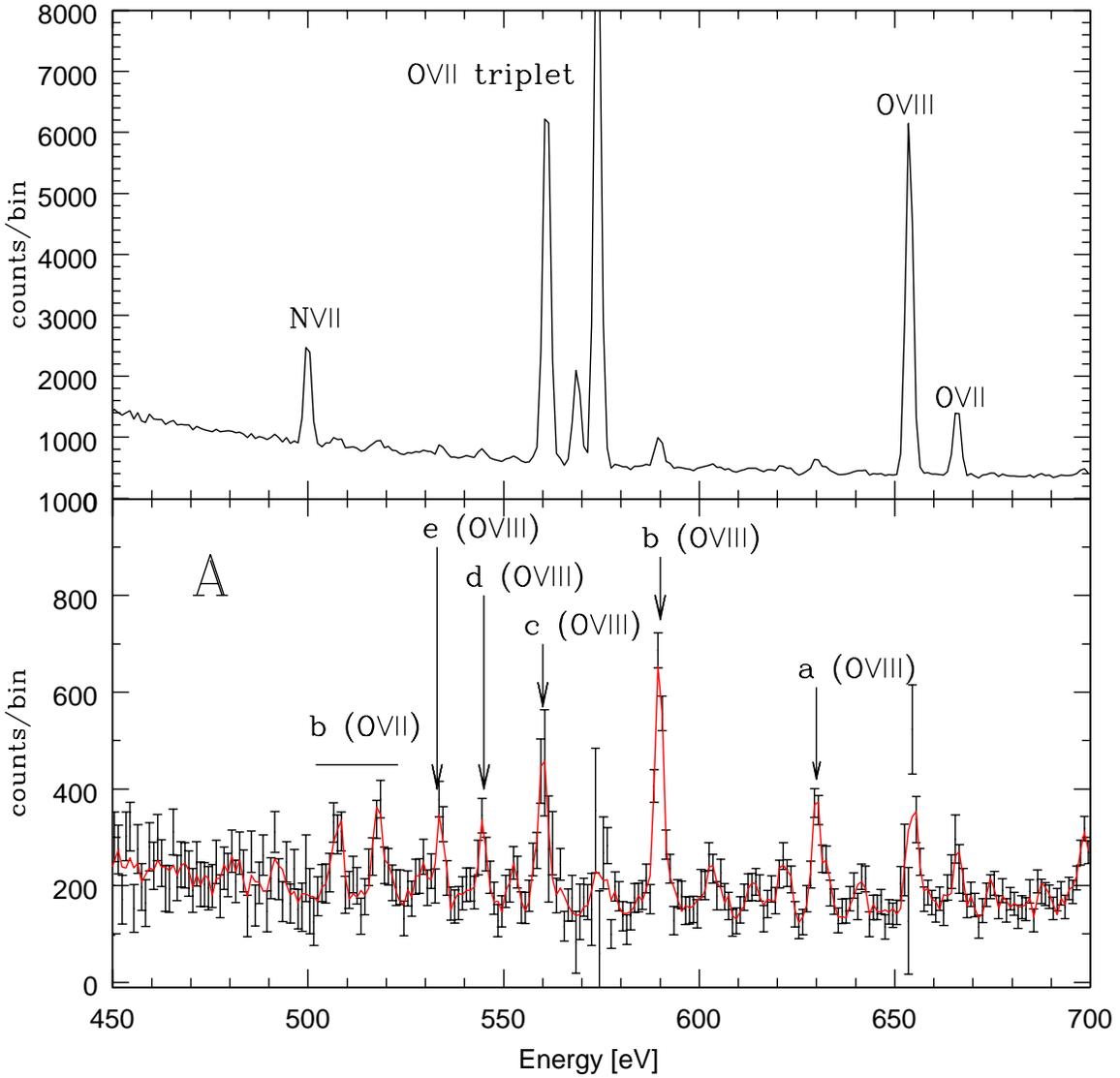}
 \end{center}
 \caption{A simulated spectrum along a line of sight. {\it Upper panel}
 shows emission lines of WHIM, the CXB and the Galactic emission. {\it
 Lower panel}: The spectrum of WHIM after the CXB and the Galactic
 emission are subtracted. The metallicity model I is
 adopted.\label{fig:spec_subtracted}}
\end{figure}

With those in mind, we perform two simulated observations according to
the above two strategies, respectively.  The first adopts $T_{\rm
exp}=3\times10^5$ sec and combines the $12\times12$ simulated
grids. Then the spatially resolved region corresponds to the the
field-of-view of 0.88 deg$^2$, and $S_{\rm eff}\Omega T_{\rm
exp}=3\times10^7$ cm$^2$ deg$^2$ sec. The second adopts $T_{\rm
exp}=10^6$ sec and combines $4\times4$ grids. Thus each spatial
resolution is now 0.098 deg$^2$, and $S_{\rm eff}\Omega T_{\rm
exp}=1.1\times10^7$ cm$^2$ deg$^2$ sec. As target regions, we select the
square regions A, B and C in Figure \ref{fig:OVIII_map} for the first
strategy and D, E, and F for the second one. Since we expect that WHIM
resides in the outskirts of galaxy clusters and/or in galaxy groups, it
is quite natural to select vicinity of rich galaxy clusters as target
regions. Actually, the regions that we select here turn out to contain a
rich galaxy cluster at redshift $z=0.04$ and several small galaxy groups
at $0.09<z<0.011$ which cannot be probed by current typical X-ray
observations. Various panels in Figure \ref{fig:OVIII_map} are displayed
to show the complementary nature of the X-ray thermal bremsstrahlung 
emission (right panels) and the oxygen emissions (left panels).
Top panels correspond to the entire lightcone for $0<z<0.3$, and the
middle and bottom panels illustrate the contribution from two regions
at $0.03<z<0.04$ and $0.09<z<0.11$.

Consider first the shallow survey.  Figure~\ref{fig:wide_los} shows the
residual emission spectra ({\it left panels}) and the redshift profiles
of gas overdensity, temperature, and O{\sc viii} surface brightness
({\it right panels}) of the three regions A, B, and C assuming model I
for metallicity. Labels in the left panels indicate the identification
of emission lines in the corresponding redshift profiles in the right
panels. Except for O{\sc vii} 665eV emission lines which locate at
$\simeq 10$eV blueward of prominent O{\sc viii} lines, most of the
emission lines without any labels are redshifted Fe\,{\sc xvii}
lines. The dashed lines in the panels of O{\sc viii} surface brightness
indicate our detection limit $3\times10^{-11}$ erg s$^{-1}$ cm$^{-2}$
sr$^{-1}$ for $S/N=10$ in the case of $S_{\rm eff}\Omega T_{\rm
exp}=3\times10^7 $ cm$^2$ deg$^2$ sec.  The filled histograms indicate
those redshift bins ($\Delta z =0.3/128$) whose surface brightness
exceeds the above limiting flux.  The region B encloses an X-ray cluster
located at $z=0.038$, and the region A contains a filamentary structure
around the cluster, as is clearly shown in the {\it middle} panels of
Figure~\ref{fig:OVIII_map}.  The region C lies close to, but is not
directly associated with, the cluster. Rather it encloses poor groups of
galaxies located at $z=0.09-0.11$ ({\it lower} panels of
Fig.~\ref{fig:OVIII_map}), which are not visible in the X-ray thermal
bremsstrahlung emission (see right panels of Fig.~\ref{fig:OVIII_map}).

The spectra along the region B exhibit a strong O{\sc viii} emission
line at $E=629$ eV, which originates from the intra-cluster medium of the
X-ray cluster at $z=0.038$ (Fig.~\ref{fig:OVIII_map}). The O{\sc vii}
triplet emission lines around $E=535-560$ eV are also ascribed to the
same cluster. The region A sweeps the outer region of the cluster which
is expected to contain less dense ($\delta\simeq20$) but hotter
($T\simeq1$ keV) IGM. In fact we do see a relatively mild O{\sc viii}
emission line at the cluster redshift. Finally in the region C we detect
a strong O{\sc viii} emission line at $E=613$ eV and O{\sc vii} triplet
lines at $E=525-540$ eV from a cluster at $z=0.065$ as well as weaker
emission lines at $E=585-605$ eV from the galaxy groups at
$z=0.08-0.12$.

Turn to next the simulated deeper survey. Figure~\ref{fig:deep_los} is
the same as Figure~\ref{fig:wide_los} but for the three regions D, E,
and F.  The detection limit for those plots is $6\times10^{-11}$ erg
s$^{-1}$ cm$^{-2}$ sr$^{-1}$ for $S/N=10$ in the case of $S_{\rm
eff}\Omega T_{\rm exp}=1.1\times10^7 $ cm$^2$ deg$^2$ sec. The region D
shows an emission line due to a substructure of a cluster at $z=0.039$,
and all the three regions exhibit the presence of galaxy groups at
$z=0.1-0.11$.  The spectra in the regions D and E show the emission
lines from the galaxy cluster and its substructure at $z=0.039$.  We
note that the emission line in the region D corresponding to the
$z=0.039$ structure is stronger than the counterpart in the region E,
although the temperature of the region D at $z=0.039$ is $\simeq
7\times10^6$ K and in fact lower than that of the region E ($\simeq
2\times10^7$ K).  This is because the emissivity of O{\sc viii}
decreases as the temperature exceeds $T \approx 3\times10^6$ K, and
clearly demonstrates that the oxygen lines are more sensitive to the
presence of the WHIM than that of the higher temperature gas associated
with intra-cluster gas. The comparison of Figures~\ref{fig:wide_los} and
\ref{fig:deep_los} with Figure~\ref{fig:OVIII_map} indeed reveals that
the WHIM that evades the detection in the X-ray is indeed detectable in
its emission features in both shallow and deep surveys that we simulate
here.

So far we show the results only for the metallicity model I.  In order
to show the dependence on the assumption of metallicity,
Figure~\ref{fig:spec_metallicity} compares the spectra of the region D
for the four metallicity models described in the last section.  Since
the emission lines in this region come from the WHIM at low redshifts,
the metallicity evolution hardly changes their intensities.  On the
other hand, the model assumption on the density dependence significantly
changes the result.  Among the four models that we consider, model IV
predicts the strongest emission lines, especially of O{\sc viii} at
$E=590$ eV and O{\sc vii} triplet at $E=500-520$ eV, while the other
three models predict very similar emission intensities.  This is because
model IV assumes the metallicity larger than $0.2Z_{\odot}$ for high
density gas particles with $\rho>10^3\bar{\rho}_{\rm b}$, which mainly
reside in the intra-cluster medium. The emission lines from mildly dense
regions including the O{\sc viii} line in $E=630$ eV, on the other hand,
are not sensitive to the metallicity model.

Figure~\ref{fig:scatter1} plots the distribution of temperatures and
densities for all the 0.098 deg$^2$ regions over the entire survey of
$5^\circ\times5^\circ$. The different symbols indicate the range of
surface brightness of emission lines for O{\sc viii} 653eV ({\it left
panels}) and O{\sc vii} 561eV ({\it right panels}). Top to bottom panels
correspond to the four different metallicity models.  Note that the
temperature and density of each region are estimated by smoothing over
the fixed {\it angular} scale (0.098 deg$^2$), and therefore the
corresponding physical smoothing lengths $R_s$ are different depending
on the redshift of each region; $R_s=1.6h^{-1}$Mpc,
$3.1h^{-1}$Mpc and $4.6h^{-1}$Mpc at $z=0.1$, 0.2 and 0.3,
respectively. Therefore, the actual density is $\sim 10$ times higher
than the smoothed density.

The O{\sc viii} emission lines from regions with $S>6\times10^{-11}$ erg
s$^{-1}$ cm$^{-2}$ sr$^{-1}$, for instance, probe the baryonic matter
with $T=10^{6}-10^{7.5}$ K and overdensity of
$\rho/\bar{\rho}=10^{0.5}-10^2$ almost independently of the metallicity
models.  The O{\sc vii} 561eV emission lines, on the other hand, are
sensitive to the WHIM with $T=10^{6.5}-10^{7.5}$K and
$\rho/\bar{\rho}=10-10^2$, relatively higher temperature and higher
density regions than those probed by O{\sc viii}.  In the temperature
range of $T=10^{6.5-7}$ K, the emissivity of O{\sc viii} is higher than
that of O{\sc vii}. Thus one expects that when the O{\sc vii} triplet
emission lines are detected, the corresponding O{\sc viii} emission line
should also show up at the same redshift. Then one may estimate of the
WHIM temperature using the line intensity ratios between O{\sc viii} and
O{\sc vii} as long as the WHIM is well approximated as a single
temperature structure. In reality, the WHIM over each smoothed region
may be in a multi-temperature phase, and in this case the temperature
estimation becomes more complicated since the O{\sc vii} emissivities
have their peaks around $T=2\sim3\times10^{6}$ K while the O{\sc viii}
emissivity is larger than those of O{\sc vii} lines at $T>3\times10^6$
K. Finally because of the rapid decrease of the emissivity of O{\sc vii}
and O{\sc viii} below 10$^6$ K, it is unlikely that one may detect WHIM
with $T < 10^6$ K through the oxygen emission lines.

\section{Implications and discussion}

In this paper, we have examined in detail the detectability of WHIM
through O{\sc viii} and O{\sc vii} emission lines using a cosmological
hydrodynamic simulation of intergalactic medium. Assuming a detector
which has a large throughput $S_{\rm eff}\Omega=10^2$ cm$^2$ deg$^2$ and
a high energy resolution $\Delta E = 2$ eV, we have presented simulated
spectra of WHIM in the collisional ionization equilibrium at soft X-ray
band $E=500-700$ eV.  Since the metallicity distribution in
intergalactic space is quite uncertain, we adopted four simple and
phenomenological models.  In all the models we found that within the
exposure time of $T_{\rm exp}=10^{5-6}$ sec our proposed detector can
reliably identify O{\sc viii} emission lines of WHIM with $T=10^{6-7}$ K
and $\delta=10^{0.5-2}$, and O{\sc vii} emission lines of WHIM with
$T=10^{6.5-7}$ K and $\delta=10^{1-2}$.  The WHIM in these temperature
and density ranges cannot be detected with the current X-ray
observations except for the oxygen absorption features toward bright
QSOs (e.g., \cite{Nicastro2002, Fang2002b, Simcoe2002}).

\begin{figure}[tbp]
 \leavevmode
 \begin{center}
  \FigureFile(65mm,65mm){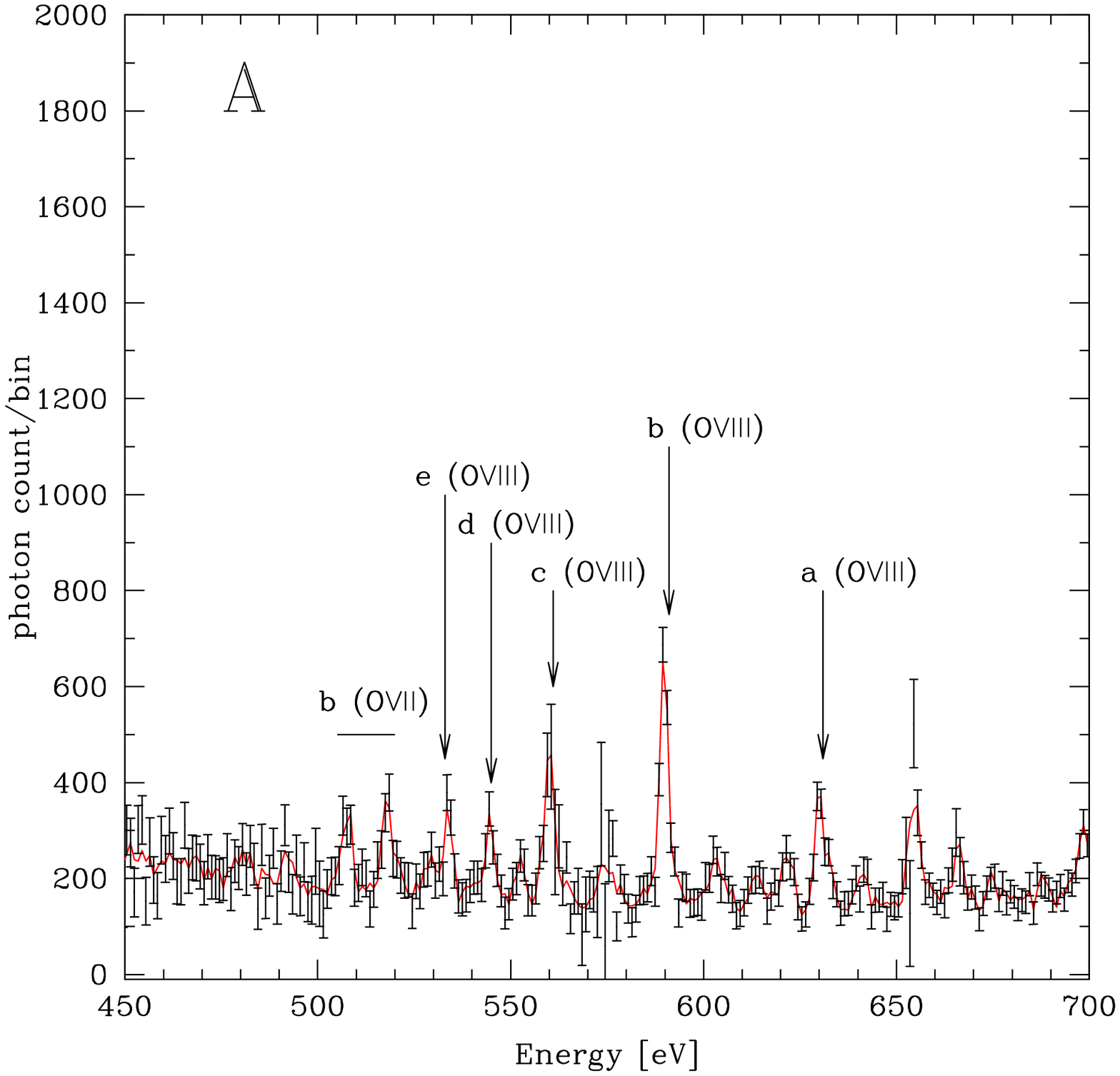}  \FigureFile(65mm,65mm){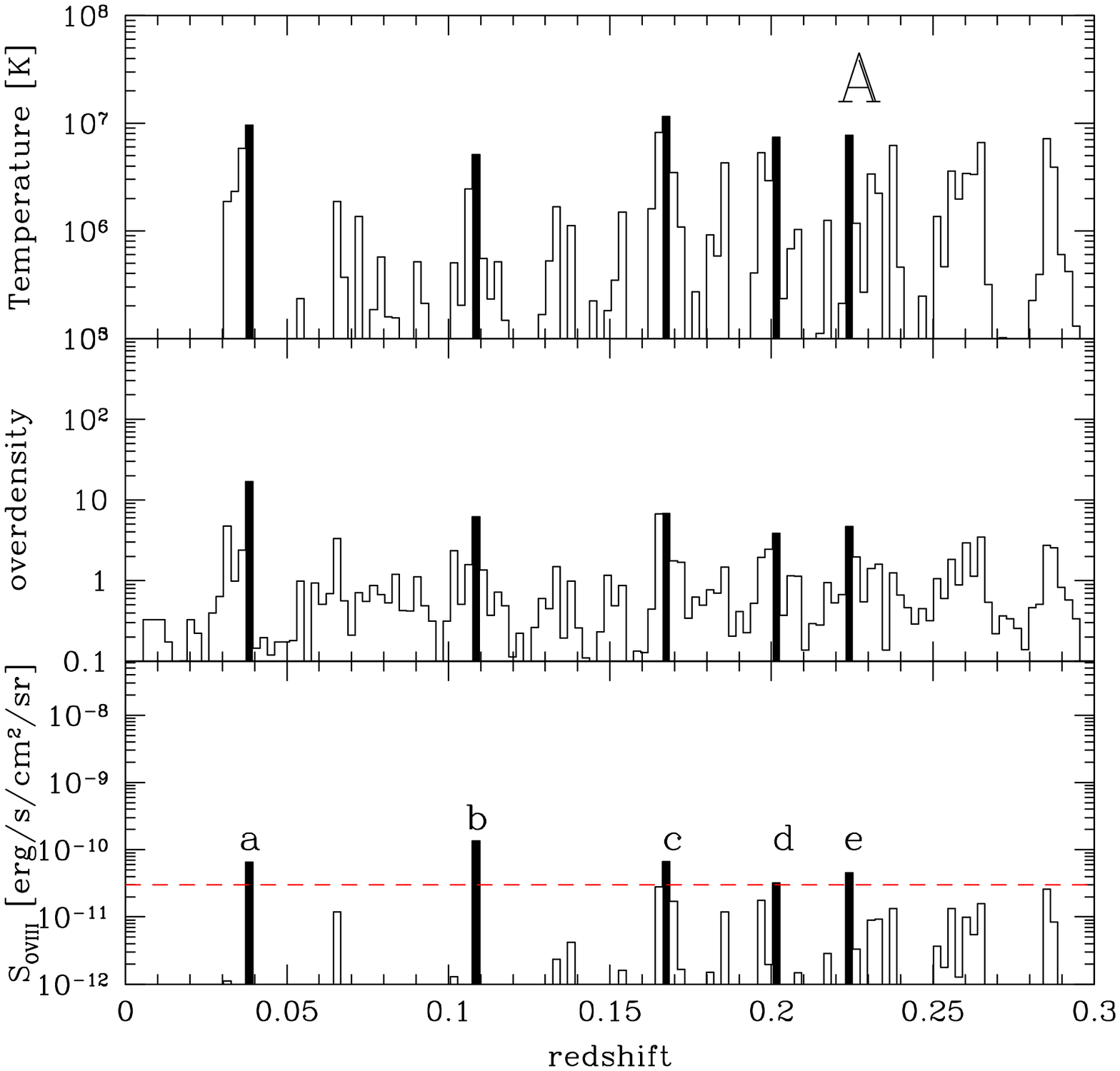} 
  \FigureFile(65mm,65mm){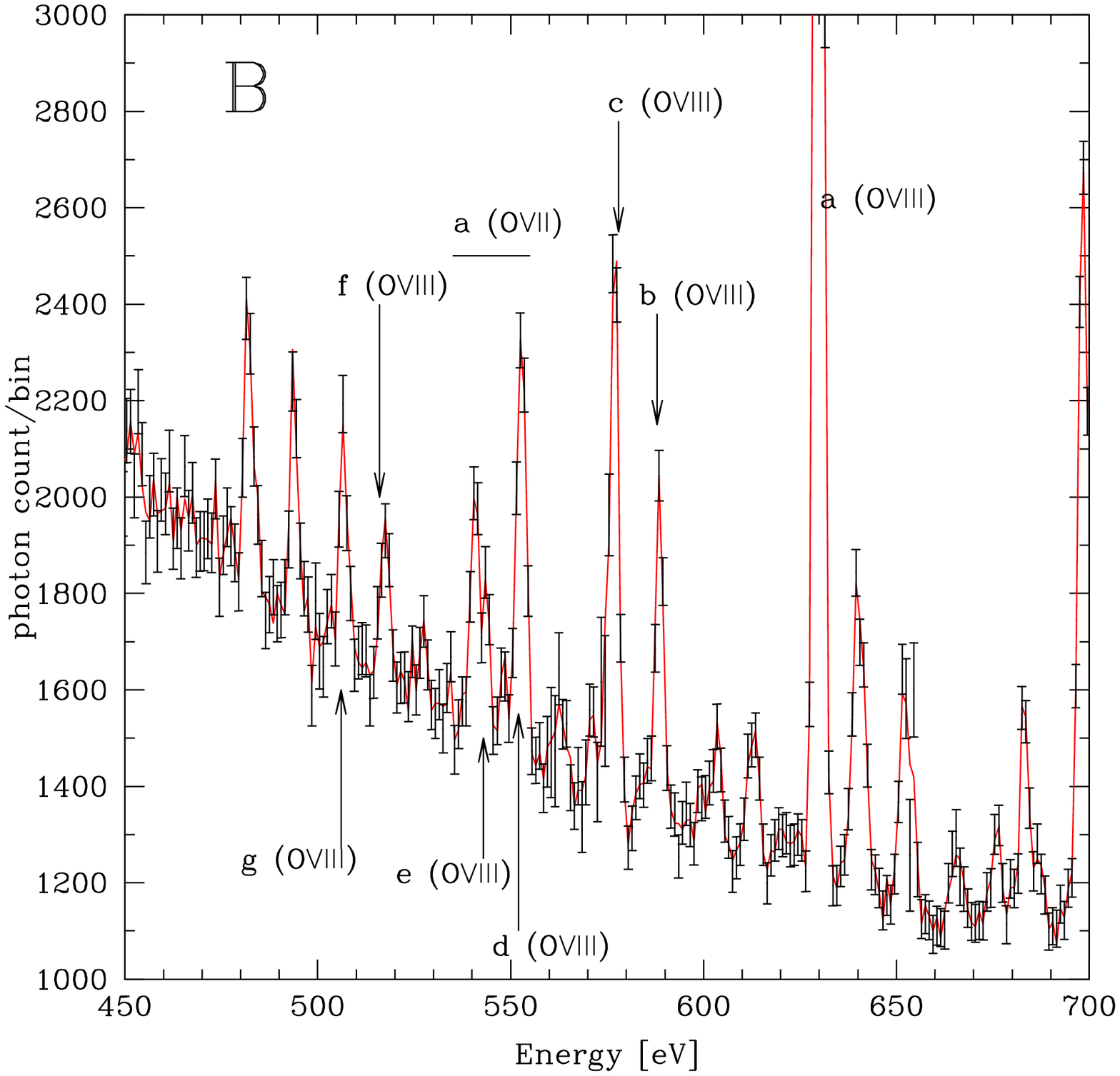}  \FigureFile(65mm,65mm){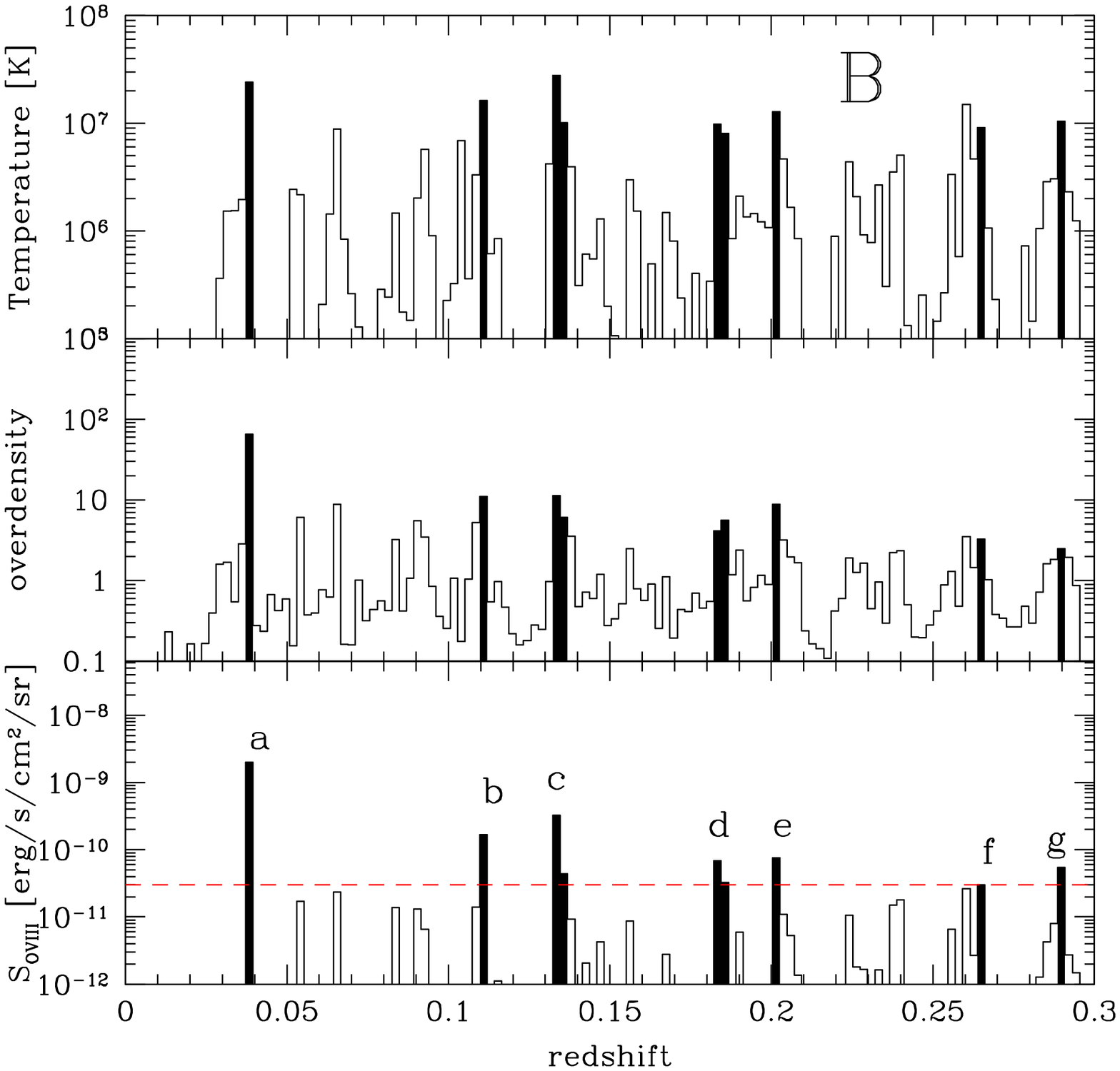} 
  \FigureFile(65mm,65mm){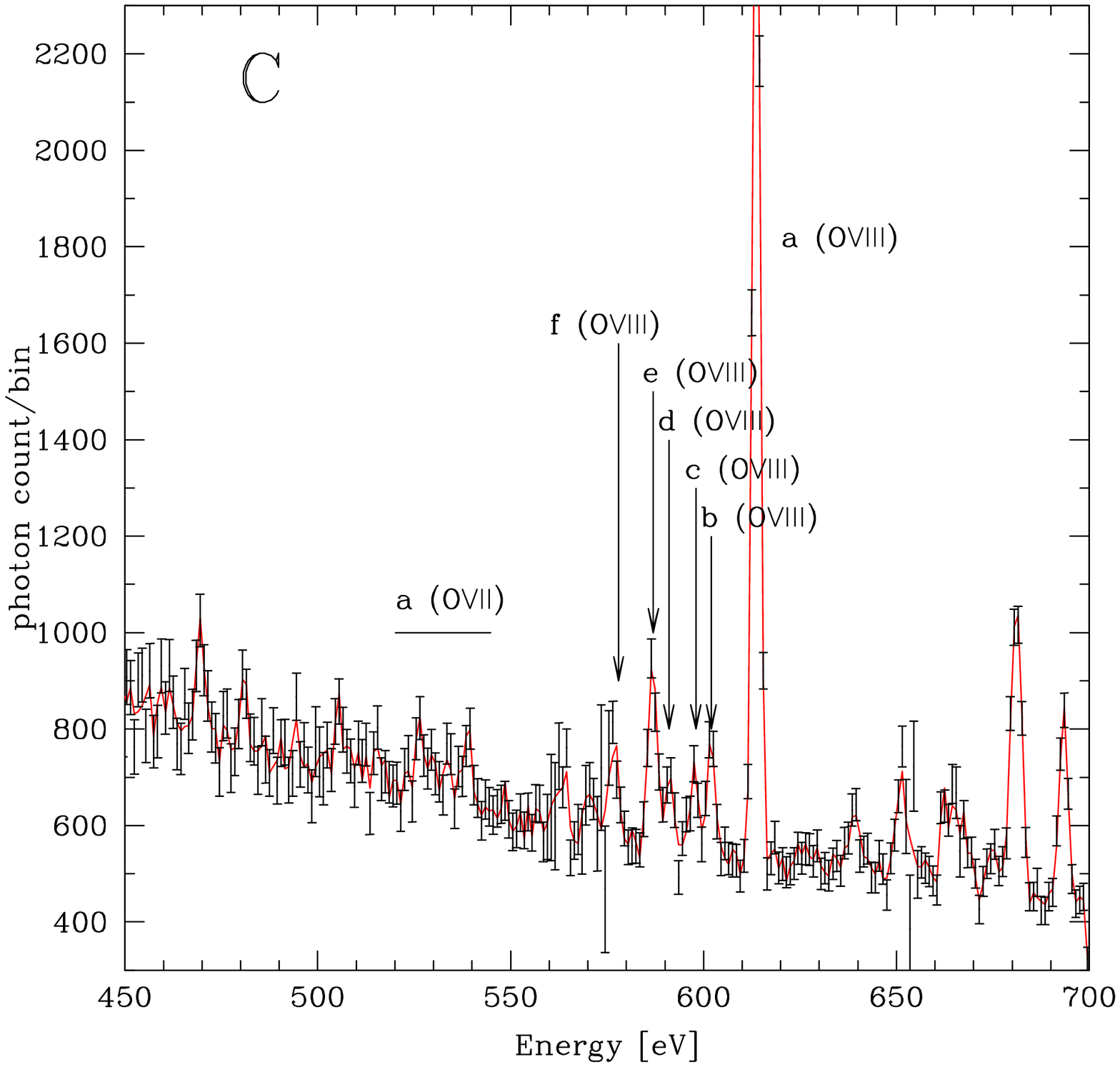}  \FigureFile(65mm,65mm){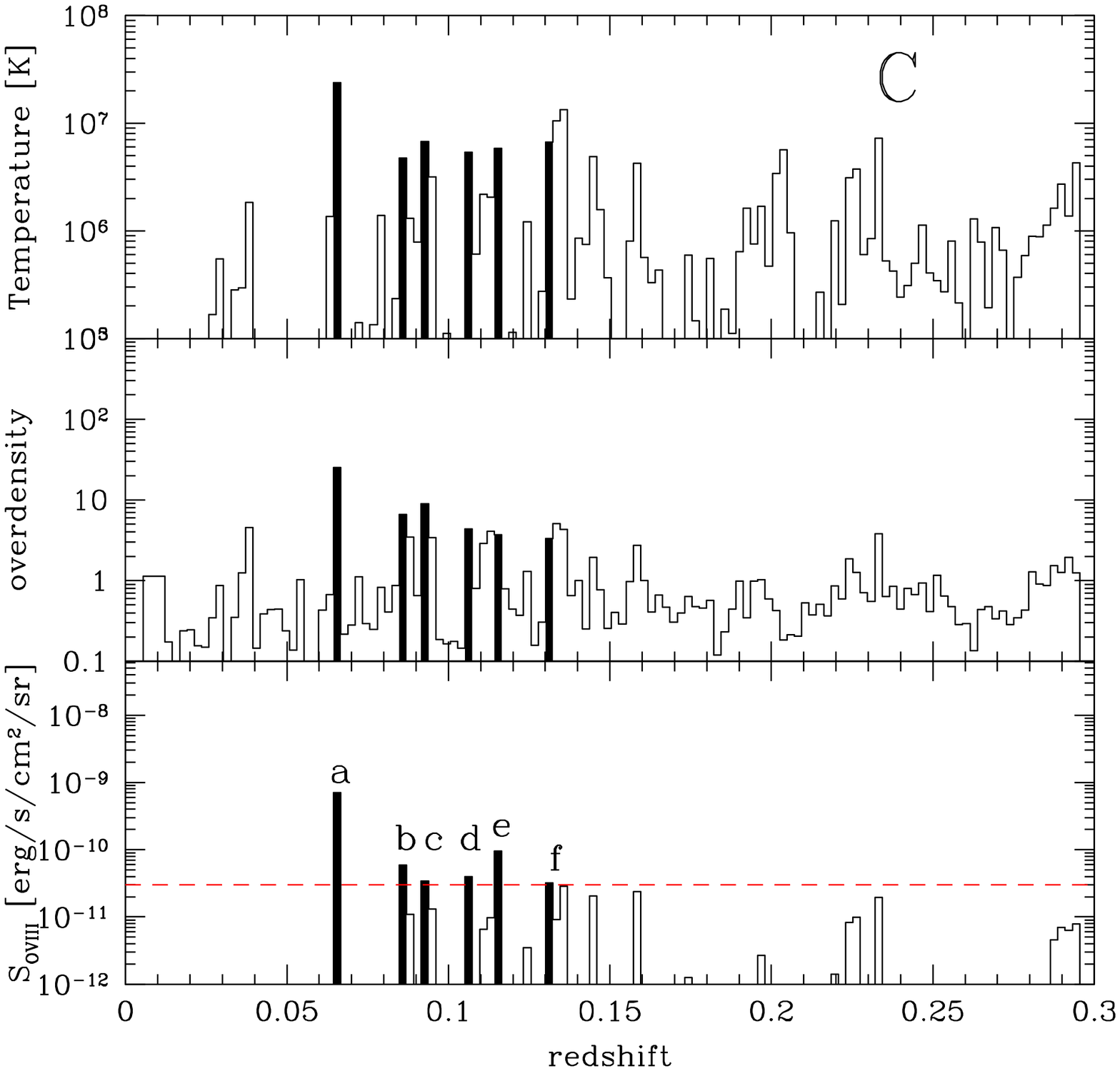} 
 \end{center}
 \caption{The simulated spectra and redshift profiles of gas density,
 temperature and O{\sc viii} surface brightness along the three
 line-of-sights marked in Figure~\ref{fig:OVIII_map}, A ({\it upper}), B
 ({\it middle}), C ({\it lower}).\label{fig:wide_los}}
\end{figure}

\begin{figure}[tbp]
 \leavevmode
 \begin{center}
  \FigureFile(65mm,65mm){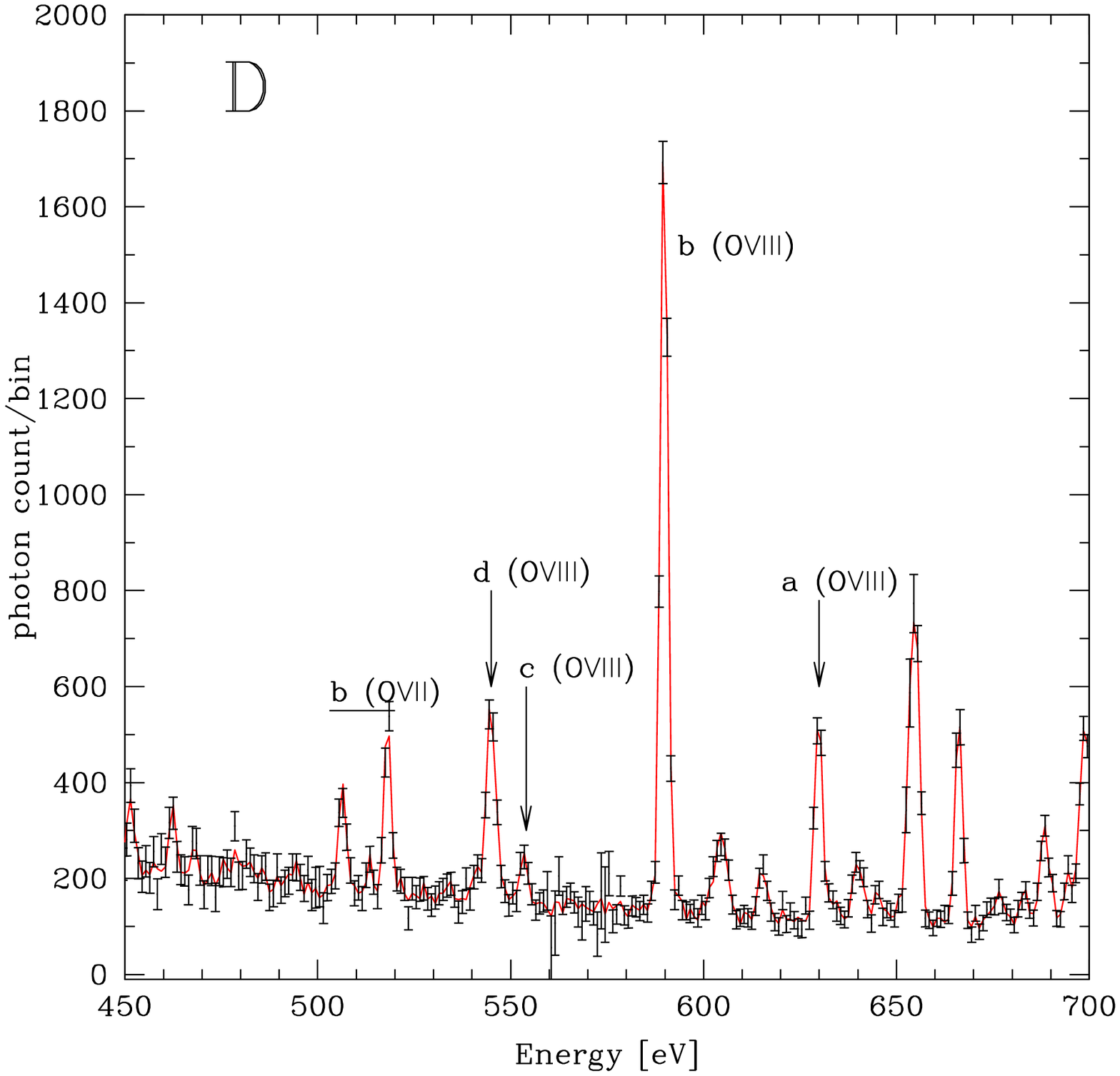} \FigureFile(65mm,65mm){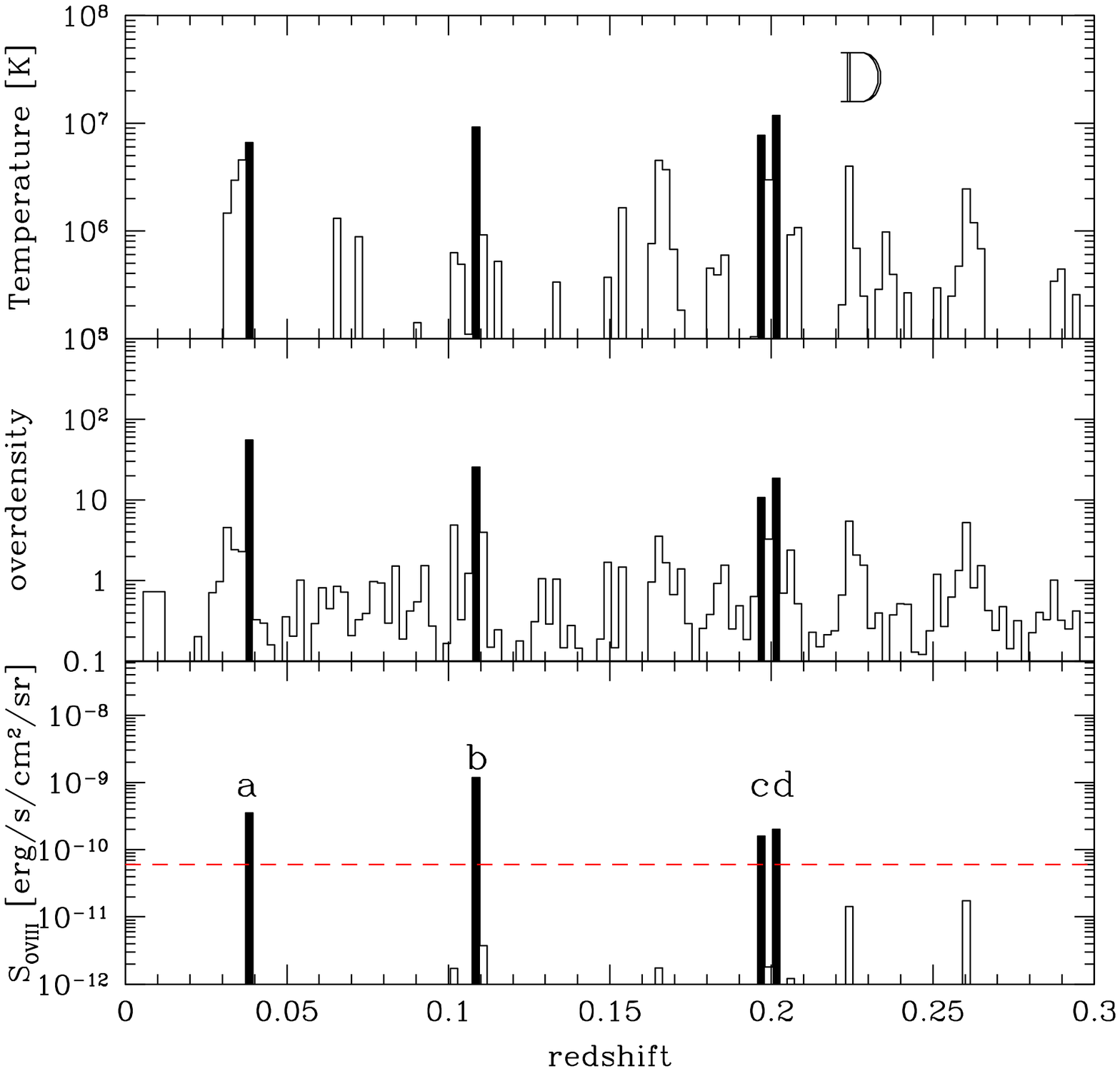} 
  \FigureFile(65mm,65mm){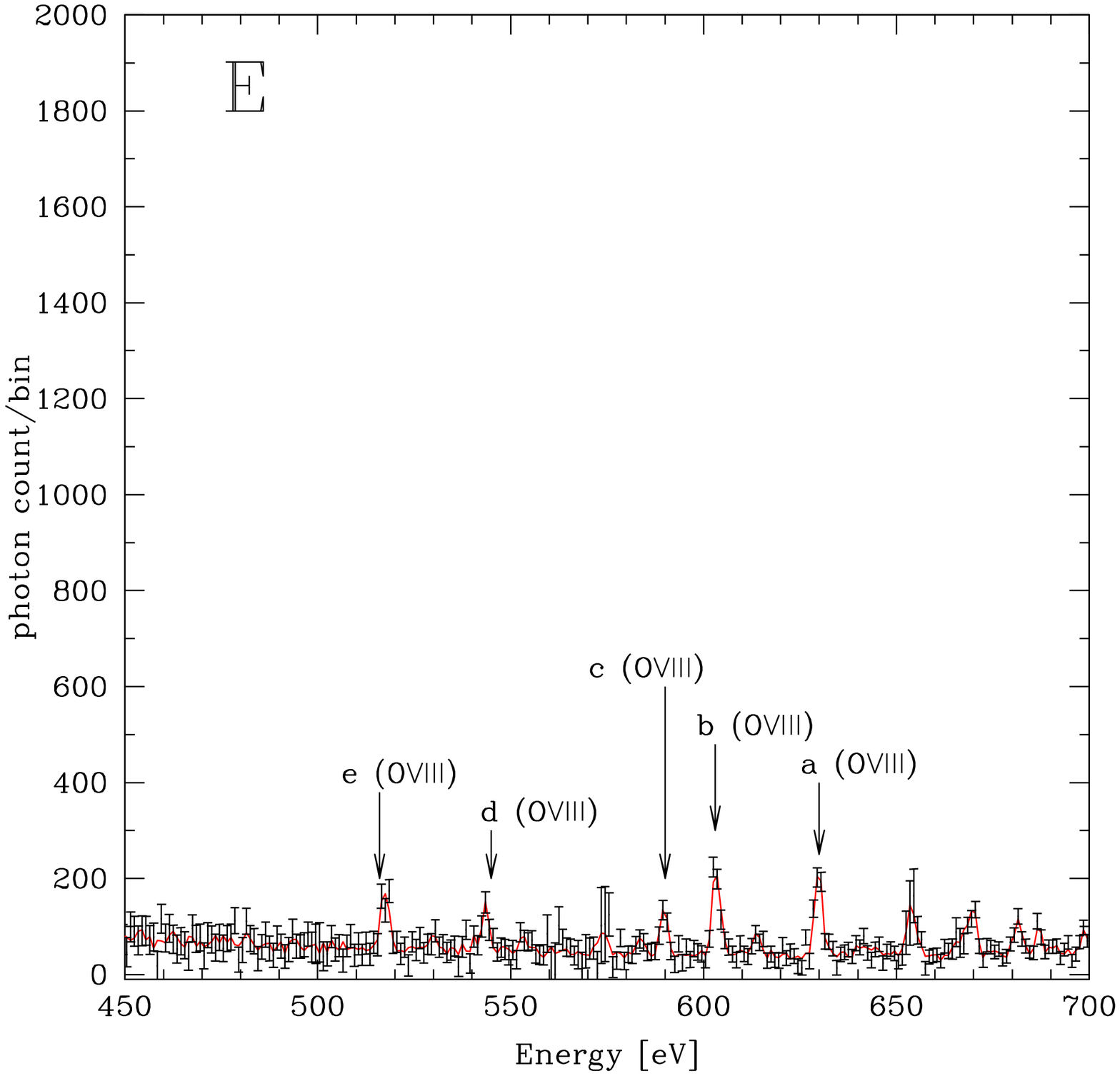} \FigureFile(65mm,65mm){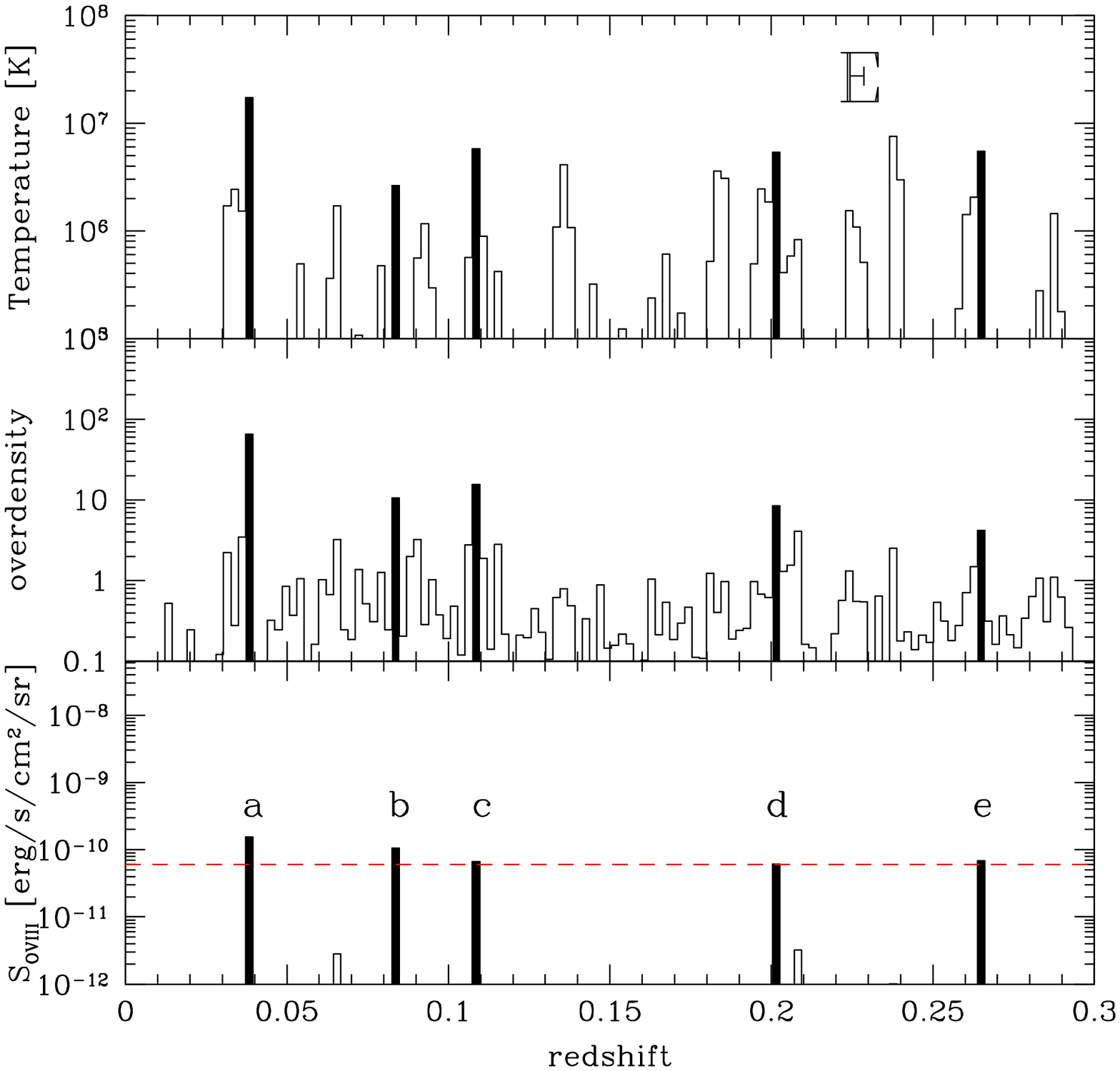} 
  \FigureFile(65mm,65mm){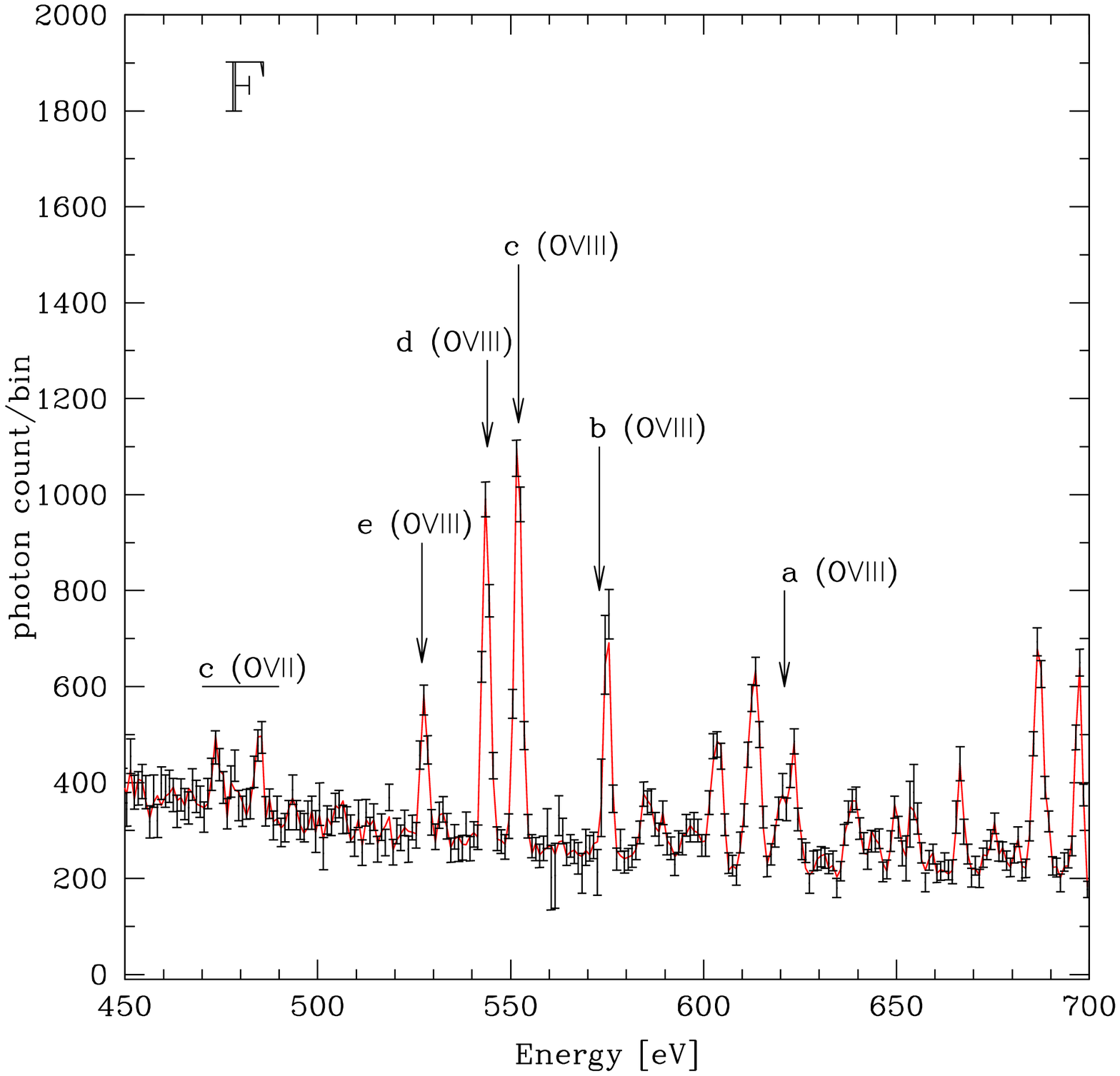} \FigureFile(65mm,65mm){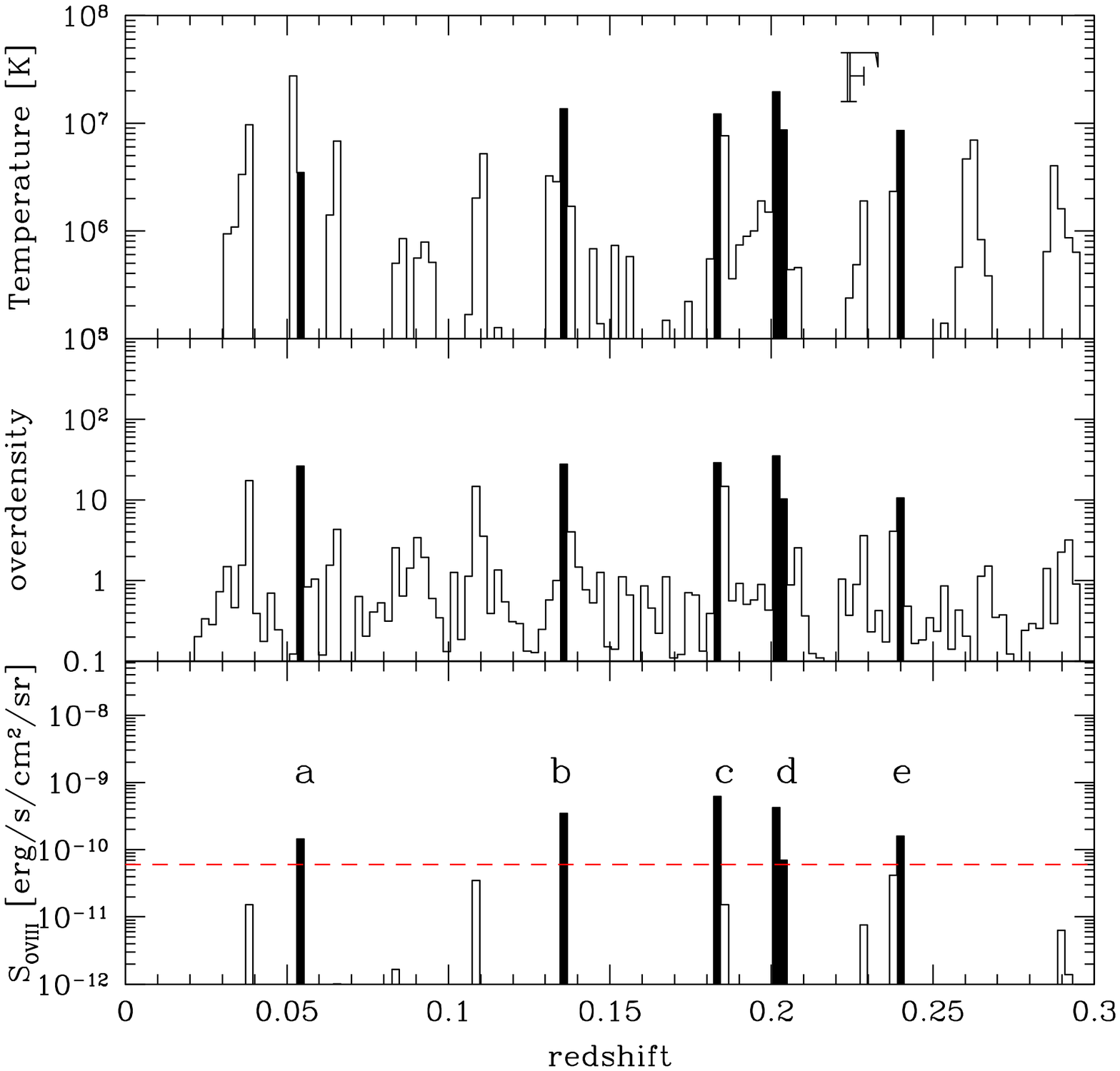} 
 \end{center}
 \caption{The simulated spectra and redshift profiles of gas density,
 temperature and O{\sc viii} surface brightness along the three
 line-of-sights marked in Figure~\ref{fig:OVIII_map}, D ({\it upper}), E
 ({\it middle}), F ({\it lower}).\label{fig:deep_los}}
\end{figure}

\begin{figure}[tbp]
 \leavevmode
 \begin{center}
  \FigureFile(160mm,160mm){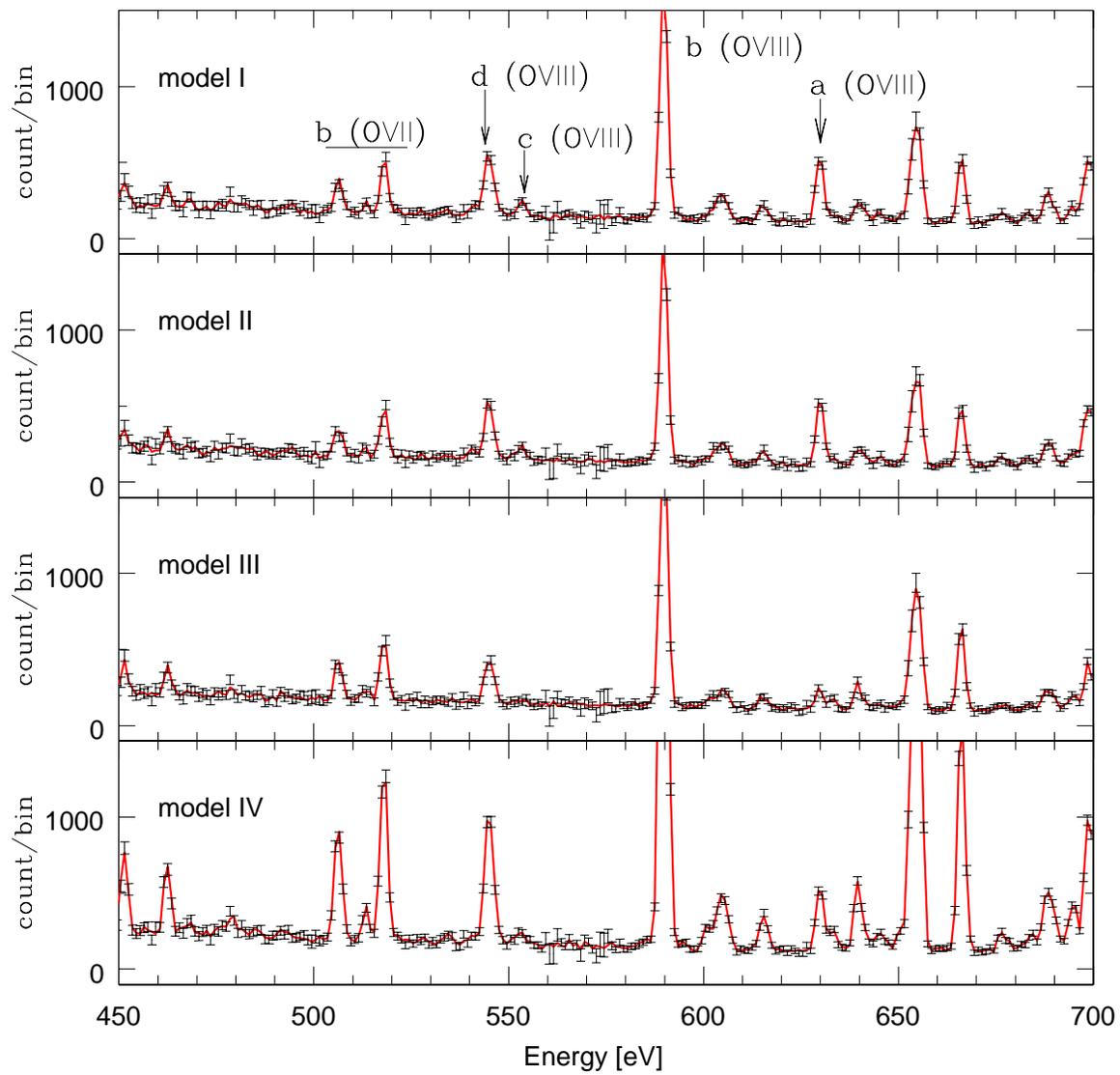}
 \end{center}
 \caption{The simulated spectra along the region D for the four
 metallicity models. \label{fig:spec_metallicity}}
\end{figure}

\begin{figure}[tbp]
 \leavevmode
 \begin{center}
  \FigureFile(115mm,115mm){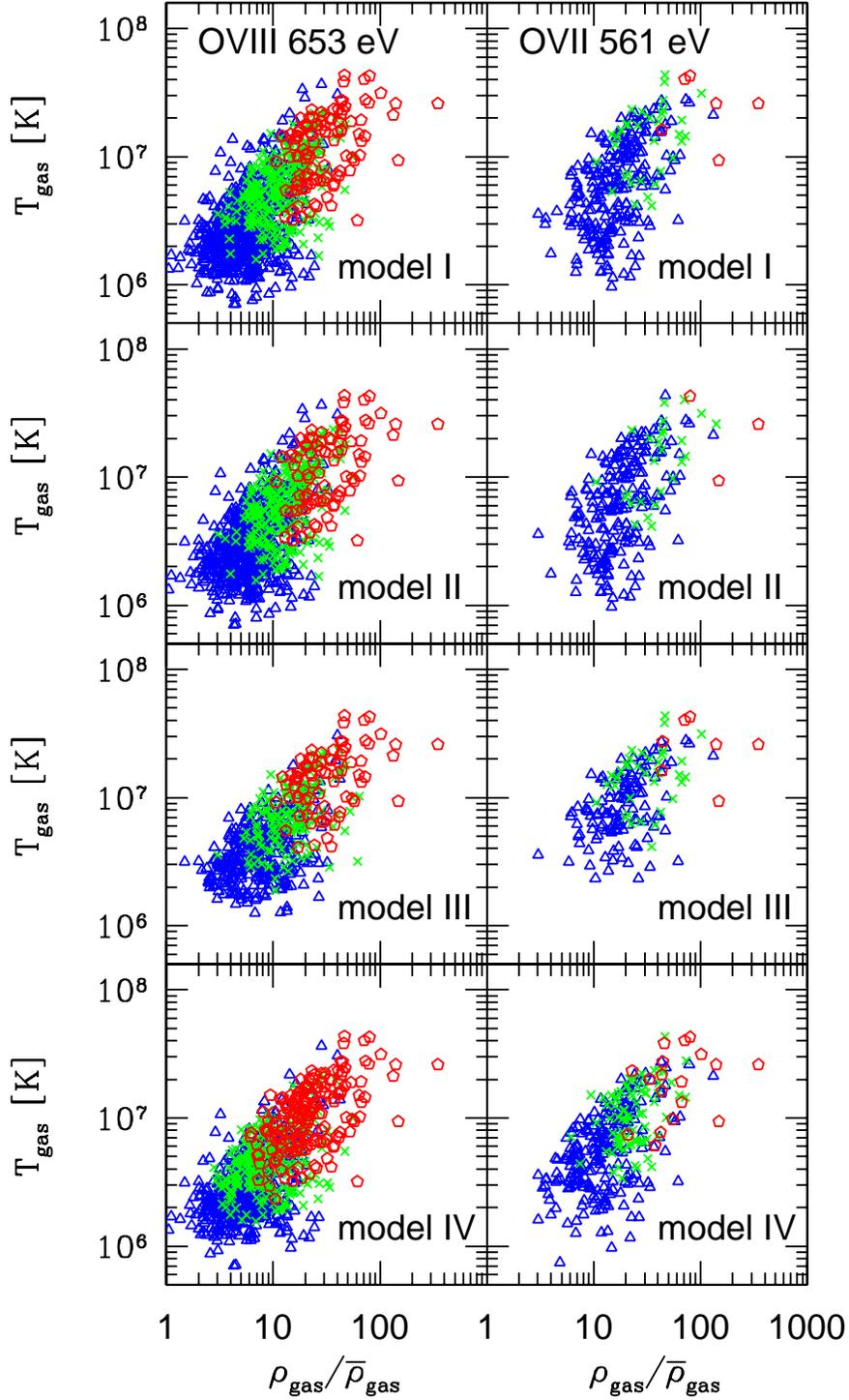} 
 \end{center}
 \caption{The distribution of baryons on the $\rho/\bar{\rho}$\,--$T$
plane for the four metallicity models.  Different symbols indicate the
three ranges of surface brightness of the emission lines; Red pentagons,
green crosses and blue triangles correspond to regions with surface
brightness $S \mbox{[erg/s/cm$^2$/sr]} >3\times10^{-10}$,
$3\times10^{-10} > S \mbox{[erg/s/cm$^2$/sr]} > 6\times10^{-11} $ and
$6\times10^{-11} > S \mbox{[erg/s/cm$^2$/sr]} > 10^{-11}$, respectively,
for the O{\sc viii} 653eV ({\it left panels}) and O{\sc vii} ({\it right
panels}) line emissions.  \label{fig:scatter1}}

\end{figure}

Let us remark three important issues related to the detectability of the
WHIM. While they are beyond the scope of the present paper, we are
currently working on those, and hope to report the results elsewhere in
due course.
\begin{enumerate}
 \item Our analysis has assumed that the WHIM is in the collisional
ionization equilibrium.  Strictly speaking, however, this assumption
may not be justified in most regions of the WHIM, since the
recombination time-scale is longer than the age of the universe. For
example, in the region with $n_{\rm e}=10^{-5}$ cm$^{-3}$ and
$T=10^6$K, the recombination time-scale of H{\sc ii} is
$3\times10^{11}$ yr, while the time-scale of the collisional
ionization is much shorter $\simeq 10^5$ yr. Thus, WHIM is thought to
be in over-ionized state, and emission lines can be produced by
recombination and subsequent cascades in contrast to the collisional
ionization equilibrium state. Unfortunately, it is not clear if the
non-equilibrium effect increases or decreases our present estimate of
the Oxygen line emission without accurate numerical
calculation. Therefore, we plan to perform a proper treatment of the
non-equilibrium ionization evolution to determine the line intensities
of O{\sc vii} and O{\sc viii}.
 \item 
Furthermore \citet{Churazov2001} pointed out that the resonant line
scattering of the CXB photons by O{\sc viii} and O{\sc vii} ions can
exceed the thermal emission in the relatively low-density WHIM with
$\delta\simeq10$ and $T\simeq10^5-10^6$ K.  Since the thermal emission
lines that we computed mostly come from the higher density and higher
temperature regions ($\delta\gtrsim10^2$ and $T>10^6$ K), the effect of
the resonant line scattering may not be dominant. Nevertheless it should
tend to increase the overall detectability of WHIM.
 \item The ability of an unambiguous identification of O{\sc vii} and
O{\sc viii} lines in the observed spectra is an important issue that
we have to address. If both O{\sc viii} and O{\sc vii} lines at the
same redshift are detected simultaneously, the identification is
relatively easy.  If one detects O{\sc viii} lines but no
corresponding O{\sc vii} counterparts, one has to worry about the
possibility for other contaminating lines such as Fe {\sc xvii} lines
and O{\sc viii} Ly-$\beta$ line. If we have very hot ($>10^7$K) at
redshift $z>0.1$, Fe {\sc xvii} lines show up at $E<650$eV and
contaminate O{\sc viii} lines at low redshift. Actually,
Figures~\ref{fig:wide_los} and \ref{fig:deep_los} show emission lines
of Fe\,{\sc xvii} in our simulated spectra for some regions, which are
significant contaminations against the proper identification of WHIM
emission line systems. Another important issue is how to estimate the
temperature and density of the WHIM from the O{\sc viii} and O{\sc
vii} lines. These are crucial in discussing the cosmological
implications of the WHIM signature from the given survey area.
\end{enumerate}

Finally we would like to emphasize that a dedicated X-ray survey mission
with the capabilities assumed in the present simulation is now under
serious consideration. The size and mass of the satellite will be as
small as (1.5 m)$^3$ and less than 300 kg. The scientific payloads will
incorporate several advanced technologies that have been developed for
the past and future Japanese X-ray astronomy missions.  The basic
structure of the X-ray telescope is a thin-foil multi-mirror type
already used for ASCA and Astro -- E II.  The major innovation in the
design optimized for the dedicated WHIM survey is the use of 4
reflection telescope, i.e., X-rays are reflected by 4 conical-shape
mirrors. This technique achieves a substantial reduction of the focal
length down to 60--70 cm, by a factor of $\gtrsim 6$ relative to that of
the 2 reflection type. Thus the extremely compact satellite becomes
possible. A preliminary study shows that our requirement, $S\Omega
\gtrsim 100$ cm$^2$deg$^2$, can be achieved in the energy band 0.5--0.7
keV with the mirror outer diameter of 36 cm. The focal plane instrument
will be an array of TES micro-calorimeters, cooled by mechanical coolers
and adiabatic demagnetization refrigerators. The energy resolution is 2
eV and the sensitive area of the detector is about 1.5 cm $\times 1.5$
cm consisting of even up to 1024 pixels. Such a detector can cover
$1^\circ \times 1^\circ$ field if the focal length is less than 86
cm. The current status of the detector development will be reported
elsewhere. We expect that within 2--3 yrs the required techniques for
the survey of WHIM will be quite feasible, and hope that the small
satellite will be launched before 2010.

\bigskip 

We thank an anonymous referee for several constructive comments.  We
also thank K.Masai and S.Sasaki for helpful discussions. Numerical
computations presented in this paper were carried out at ADAC (the
Astronomical Data Analysis Center) of the National Astronomical
Observatory, Japan (project ID: mky05a).  This research was supported in
part by the Grants-in-Aid by Monbu-Kagakusho, Japan (07CE2002, 12304009,
12440067, 12640231).

\newpage
\onecolumn
\begin{table}
 \begin{center}
  \caption{Specification and Detection limit of emission lines for
  various future and current X-ray satellites.\label{tab:detector}}
  \begin{tabular}{lcccccc}
   \hline\hline
   \multicolumn{1}{c}{satellite} & year & $S_{\rm eff}$ [cm$^2$] 
   & $S_{\rm eff}\Omega$ [cm$^2$deg$^2$] & $\Delta E$ [eV] & $f_{\rm lim}$
   [erg/s/cm$^2$/sr] \\
   \hline
   Chandra ACIS-S3 & 1999 & 600         & 12    & 80   & $10^{-9}$\\
   XMM-Newton EPIC-pn & 1999 & 1200 & 100   & 80   & $3\times10^{-10}$\\
   Astro-E II  XIS & 2005 & 90$\times$4 & 36    & 80   & $6\times10^{-10}$\\
   Astro-E II  XRS & 2005 & 90          & 0.23  &  6   & $2\times10^{-8}$\\
   Const-X SXT+TES & mid-2010's & 3000  & 5.6   &  2   & $7\times10^{-10}$\\
   XEUS-I          & mid-2010's & 60000 & 16.7  &  2   & $2.5\times10^{-10}$\\
   our model detector & $\sim 2010$ & (100$\sim$200)  & 100   &  2 & $6\times10^{-11}$\\
   \hline
  \end{tabular}
 \end{center}
\end{table}

\end{document}